\documentclass[12pt,preprint]{aastex}
\def\Lbsun{\hbox{$\thinspace L_{B\odot}$}}
\def\Mbsun{\hbox{$\thinspace M_{B\odot}$}}
\newcommand{\gapprox}{{_> \atop{^\sim}}}

\slugcomment{\hbox to 6.5in{Accepted for publication in
{\em The Astrophysical Journal} \hfil           \hfil
\hbox{LPNHE~02-02}}}    
\begin{document}
\title{The distant Type Ia supernova rate}
\author{
R.~Pain\altaffilmark{1},
S.~Fabbro\altaffilmark{1,2},
M.~Sullivan\altaffilmark{3},
R.~S.~Ellis\altaffilmark{4,5},
G.~Aldering\altaffilmark{6},
P.~Astier\altaffilmark{1},
S.~E.~Deustua\altaffilmark{6},
A.~S.~Fruchter\altaffilmark{7},
G.~Goldhaber\altaffilmark{6,8},
A.~Goobar\altaffilmark{9},
D.~E.~Groom\altaffilmark{6},
D.~Hardin\altaffilmark{1},
I.~M.~Hook\altaffilmark{10},
D.~A.~Howell\altaffilmark{6},
M.~J.~Irwin\altaffilmark{4},
A.~G.~Kim\altaffilmark{6},
M.~Y.~Kim\altaffilmark{6},
R.~A.~Knop\altaffilmark{6},
J.~C.~Lee\altaffilmark{6,11},
C.~Lidman\altaffilmark{12},
R.~G.~McMahon\altaffilmark{4},
P.~E.~Nugent\altaffilmark{6},
N.~Panagia\altaffilmark{7},
C.~R.~Pennypacker\altaffilmark{6,13},
S.~Perlmutter\altaffilmark{6,8},
P.~Ruiz-Lapuente\altaffilmark{14},
K.~Schahmaneche\altaffilmark{1},
B.~Schaefer\altaffilmark{15,16},
N.~A.~Walton\altaffilmark{4}\\[0.25cm]
{\large (The Supernova Cosmology Project)}
}
\altaffiltext{1}{LPNHE, CNRS-IN2P3 and Universit\'es Paris VI \& VII,
Paris, France}
\altaffiltext{2}{Now at IST, Lisbon, Portugal}
\altaffiltext{3}{Department of Physics, Durham University, United Kingdom}
\altaffiltext{4}{Institute of Astronomy, Cambridge University, United Kingdom}
\altaffiltext{5}{California Institute of Technology, 
Pasadena, California}
\altaffiltext{6}{E. O. Lawrence Berkeley National Laboratory,
Berkeley, California}
\altaffiltext{7}{Space Telescope Science Institute, Baltimore, Maryland}
\altaffiltext{8}{Center for Particle Astrophysics, U.C. Berkeley, 
Berkeley, California}
\altaffiltext{9}{Physics Department, University of Stockholm, Sweden}
\altaffiltext{10}{Institute for Astronomy, Royal Observatory,  
Edinburgh, United Kingdom}
\altaffiltext{11}{Now at MIT, Center for space research, 
Cambridge, Massachusetts}
\altaffiltext{12}{European Southern Observatory, La Silla, Chile}
\altaffiltext{13}{Space Sciences Laboratory, U.C. Berkeley, 
Berkeley, California}
\altaffiltext{14}{Department of Astronomy, University of Barcelona, 
Barcelona, Spain}
\altaffiltext{15}{Department of Astronomy, Yale University, 
New Haven, Connecticut}
\altaffiltext{16}{Current address: University of Texas, 
Austin, Texas}
\begin{abstract}
We present a measurement of the rate of distant Type Ia supernovae
derived using 4 large subsets of data from the Supernova Cosmology
Project. Within this fiducial sample, which surveyed about 
12 square degrees,  
thirty-eight supernovae were detected at redshifts $0.25$--$0.85$.
In a spatially-flat
cosmological model consistent with the results obtained by the
Supernova Cosmology Project, we derive a rest-frame Type Ia supernova
rate at a mean redshift $z\simeq0.55$ of
$1.53\,{^{+0.28}_{-0.25}}\,
{^{+0.32}_{-0.31}}\,10^{-4}\,h^3\,{\rm Mpc}^{-3}\,{\rm yr}^{-1}$ or
$0.58\,{^{+0.10}_{-0.09}}\,{^{+0.10}_{-0.09}}\,h^2\,{\rm SNu}$ (1
SNu = 1 supernova per century per $10^{10}$\Lbsun), where the first
uncertainty is statistical and the second includes systematic
effects. The dependence of the rate on the assumed cosmological
parameters is studied and the redshift dependence of the rate per
unit comoving volume is contrasted with local estimates in the context of
possible cosmic star formation histories and progenitor models.
\end{abstract}
\section{Introduction}
Recent observational efforts to detect high-redshift 
supernovae (SNe) have clearly demonstrated their value as
cosmological probes. For the primary purpose of constraining the
cosmic expansion history, the Supernova Cosmology Project (SCP)
developed a scheduled search-and-follow-up technique 
that allows the systematic, on-demand discovery and follow up
of ``batches'' of high-redshift SNe
\cite[]{Perlmutter:1995a}. 
Such batch discoveries of supernovae over the following years 
have led to the
construction of two largely independent Hubble diagrams,
one by the SCP
\cite[]{Perlmutter:1997a, Perlmutter:1998, Perlmutter:1999}
and one by the High-Z Supernovae Search Team
\cite[]{Garnavich:1998, Schmidt:1998, Riess:1998}, which 
both indicate significant, non-zero cosmological constant.

The batch discovery technique also provides well-controlled search
conditions that make it possible to measure the 
{\em rate of occurrence} of distant SNe.
In 
Pain et~al (1996; hereafter Paper~I), we presented the 
first such measurement using this technique. 
The distant supernova rate, and its comparison with the nearby 
supernova rate, can provide 
a diagnostic of the cosmic star formation
history and metal enrichment at high-redshift, as well as a
better understanding of possible SNe\,Ia progenitor models
\citep{Madau:1998,Yungelson:1998}. 
Obtaining a broader understanding 
of the nature and origin of high-redshift SNe 
will further improve and refine our use of supernovae as 
cosmological probes.

The local SNe\,Ia rate has recently been reported for two samples, one with
$z\simeq0.01$ \citep{Cappellaro:1999} based on visual and
photographic plates searches, and another at $z\simeq0.1$
\citep{Hardin:2000} based on CCD searches. 
In Paper~I, we presented 
the SN\,Ia rate at intermediate redshift ($z$)
($z\simeq0.4$) using three SNe\,Ia discovered with the 2.5-m Isaac
Newton Telescope (INT). In this current paper, we report a refined
measurement based on an enlarged sample of 38 SNe\,Ia, spanning
the redshift interval $0.25$--$0.85$, discovered over the course
of four observing runs at the Cerro Tololo 4-m telescope. The new
sample allows us, for the first time, to place constraints on the
important question of possible evolution in the rate.

The method we adopt to calculate the SN rate is described in
detail in Paper~I, and contains two components. The first is the
estimation of the SN detection efficiency and hence the ``control
time'' (the effective time during which the survey is sensitive to
a Type Ia event). We have studied our detection efficiency as a
function of magnitude and supernova-to-host-galaxy surface
brightness ratio using Monte-Carlo techniques. The second part
estimates the comoving volume and total stellar luminosity to which our SNe
survey is sensitive. We have computed the total galaxy luminosity
from galaxy counts estimated from the Canada-France Redshift
Survey (CFRS) and, independently, from recent parameterizations of
the type-dependent field galaxy luminosity function and its
redshift evolution. In combination, both aspects then yield an
accurate determination of the SN\,Ia rate at a mean redshift of
$z\simeq0.55$.

A plan of the paper follows. In $\S$2 we discuss the new SN
dataset and in $\S$3 introduce our methodologies for estimating
the control time and detection efficiencies. We reach
significantly fainter detection limits compared to those
of Paper I. In $\S$4 we introduce the formalism for
determining the survey comoving volume and in $\S$5 various ways for
estimating the accessible total stellar luminosity. This allows us
to estimate the intermediate redshift SN rate in SNu (1 SNu = 1
supernova per century per $10^{10}$\Lbsun). We discuss the various
components of the uncertainties, statistical and systematic, in
$\S$6 and interpret our results in the context of local estimates
and cosmic star formation histories in $\S$7.

\section{The Data Sets}

For this analysis, we have studied 4 independent 
datasets of roughly equal size, totaling 
219 similar search fields.
These
fields were observed in November and December of 1995 (Set A), in
February and March of 1996 (Set B), in February and March of 1997
(Set C), and finally in November and December of 1997 (Set D), all
using the Cerro Tololo 4-m telescope in Chile. The data sets were
obtained as part of the search for high-redshift SNe conducted by
the SCP.
These images are suitable for a determination of the SN
rate since they were obtained under similar conditions at one
telescope, and therefore form well-defined, homogeneous sets.

Sets A and B were obtained using the $2048^2$-pixel prime-focus
CCD camera, whereas Sets C and D were obtained with the
$4\times2048^2$-pixel Big Throughput Camera (BTC,
\citet[]{Wittman:1998}).
The projected pixel size is $\simeq0.43''$ in both cases, giving an
image size of approximately $16'\times 16'$ 
(or
$4\times 16'\times16'$ 
with the BTC). Exposure
times were $2\times 600~{\rm s}$ or more in the 
Kron-Cousins $R$
filter, and the individual images reach a point-source $3\sigma$ magnitude
limit ranging from $R=22.5$\,mag to $R=24.5$\,mag.  Seeing was
typically around $1''$. The fields lie in the range
$0^h<\alpha<15^h$, $\delta>-10^{\circ}$, 
avoiding the Galactic
plane ($|b|\gapprox 30^{\circ}$). A few of the fields were
selected due to the presence of a high-redshift cluster. 
The effect of the presence of clusters in the
survey fields is taken into account in the calculation of the SN
rate (see $\S$4).

For all fields, a first-look ``reference'' image was obtained followed
by a second look ``search'' image $2$--$3$ weeks later.  The useful 
solid angle
of this dataset is defined by the overlap region of the original set of
reference images with the search images. The total useful solid angle 
covered in
this study is $\simeq12$ square degrees. The ``reference'' images were 
subtracted from the ``search'' images after convolution to match the 
seeing of the worst image and scaling in intensity. The resulting 
difference image for each field was searched for SN candidates.
Tables~1a-d give the coordinates of the fields
together with the supernova detection limit and the color excess 
$E(B-V)$ derived from \citet{Schlegel:1998}.

\paragraph*{Supernova Detection and Identification}
The original search for supernovae was performed with a view to measure 
the cosmological parameters $\Omega_{\rm M}$ and $\Omega_\Lambda$
\cite[]{Perlmutter:1999}. 
The detection of supernovae was done 
in three steps: (1) the selection of transients events detected on the 
subtraction images with a signal-to-noise ratio cut of $3.5\sigma$ and
a $15\%$ cut on the ratio of the candidate flux and 
the host galaxy aperture flux at the candidate position. The latter
cut had to be applied to remove systematics from subtraction residuals;
(2) the rejection of statistical fluctuations, cosmic rays, asteroids with  
coincidences built from the multiple images of the same field 
taken at both epochs (``reference'' and ``search'');
(3) the rejection of the remaining spurious candidates generated
by hot or dead pixels, flatfields defects or bad subtractions with  
a visual inspection of each subtraction. 

Altogether, 58 candidates passed the cuts in the original search 
and all but one were observed 
spectroscopically with the Keck Low-Resolution Imaging Spectrometer (LRIS, 
\citet{Oke:1995}). 
The one remaining candidate was not followed up spectroscopically
due to a lack of telescope time
(and was thus not included in the cosmological parameter study in 
\citet{Perlmutter:1999}). Its light curve, however, is 
consistent with that of a SN\,Ia at redshift $z\simeq0.7$.  
Of the 57 objects with spectral information, 4 were classified  
as ``non supernovae'' (QSO/AGN) and the 53 remaining retained as 
possible supernovae
\cite[]{Perlmutter:1995b, 
Perlmutter:1996, Perlmutter:1997b, Perlmutter:1997c}.

For the purpose of measuring the rate, a new search was performed on the 
same subtractions, slightly raising the signal-to-noise ratio cut (typically 
to $5\sigma$) in order to ensure good
control of the supernova detection efficiencies. 
Forty-six candidates remained at this stage 
(including the 4 ``non supernovae'') of which 5 were spectroscopically 
identified 
as ``non Ia'' (II or QSO/AGN or Ib/c) and 37 as ``possible SN Ia''.
Type II supernovae were identified by the presence of hydrogen or by their
very blue featureless spectrum, Ib/c by the absence of hydrogen and 
Si~II or S~II lines and the presence of narrow Ca~II H\&K features.
The following criteria were then used to identify the SN~Ia
\cite[]{Hook:2002}: (1)
presence of Si~II in the spectrum. For redshifts greater than
$z\sim0.5$, the Si~II~$\lambda4130$\AA\, 
was used since Si~II~$\lambda6150$\AA\, 
is beyond the spectroscopic range of {\it LRIS};  
(2) presence of S~II ``W'' feature at $\sim5500$\AA\, when detected;
(3) the large width of the $\sim4000$\AA\, Ca~II feature, characteristic of 
Type~Ia SN.    

Twenty eight candidates were identified as SN~Ia using the above criteria 
leaving only nine for which the spectra had signal-to-noise ratio too low 
to distinguish among Type~I sub-types. These 9 objects
were discovered during the first 2 runs (Set A and set B) and 
observed spectroscopically under non optimal weather conditions. 
On the contrary, all objects discovered during the 2 other runs 
(Set C and set D) were observed with good signal-to-noise ratios. 
None of these events were classified as Ib/c. 
Considering the fact that all 4 sets have roughly equal sizes 
and were searched using the same procedures, this implies that 
the contamination by non-SN~Ia in sets A and B is likely to be 
comparable, i.e. less than 10\%.
Two candidates have a E/S0 host type~\cite[]{Sullivan:2002} which is a 
strong indicator of the supernova being of Type~Ia.
Adding the facts that the light curves of these partially 
identified objects resemble a Type~Ia light curve at the observed 
redshift and that their peak 
magnitude is close to a Type~Ia peak magnitude,
we classified all 9 objects as ``probable Ia''. 
These 9 events together with the one which was not observed 
spectroscopically were therefore retained  
for the rate analysis but the possibility that one of these objects  
may not be a SN\,Ia was used to estimate the effect of possible 
misidentification of supernova type on the systematic uncertainty ($\S$6).

Altogether,
thirty-eight SNe~Ia with redshifts ranging from 0.25 to 0.85 were
retained from the fifty-eight discovered. 
Redshifts were determined from spectra of the host galaxies.
%
The properties of all 38 SNe\,Ia used in this analysis are
summarized in Table~\ref{tabsn}.

\section{Detection Efficiencies and Control Time}

The data presented here were obtained with an observing strategy 
designed to
measure the cosmological parameters $\Omega_{\rm M}$ and $\Omega_\Lambda$
by conducting a search for supernovae on the rise using a subtraction
technique. 
We followed the procedure introduced in Paper~I
to calculate the ``control time'' and detection efficiencies.

\paragraph*{Supernova Detection Efficiencies}
Detection efficiencies were determined for every search field
using Monte-Carlo simulations. A synthetic image was created for every 
field by adding simulated SNe to the search images. Reference images
were subtracted from the synthetic search images using exactly the
same software and cuts as used for the actual search, and the number of
simulated SNe that satisfied the selection criteria was determined.
The efficiency derived in this way then naturally accounts for
parts of the image that are unusable for the SN search, for
example regions saturated by bright foreground stars. Over two
hundred simulated SNe were placed on each search image, with a
range of apparent magnitude, host galaxy apparent magnitude and
location with respect to host galaxies.  Each simulated SN was
generated by scaling down and shifting a bright star, with
signal-to-noise ratio greater than 50, from the image being
studied (it was not necessary to add additional Poisson noise
because the dominant noise source is that of the sky).
The position relative to the host galaxy was chosen at
random from normal distributions with $\sigma$ equal to the half
width at half maximum of the galaxy independently on both $x$ and $y$
axis. The shift of the scaled bright star relative to the host galaxy was
constrained to be an integral number of pixels in order to maintain 
the pixelized point spread function.

\begin{figure}
\figurenum{1}
\plotone{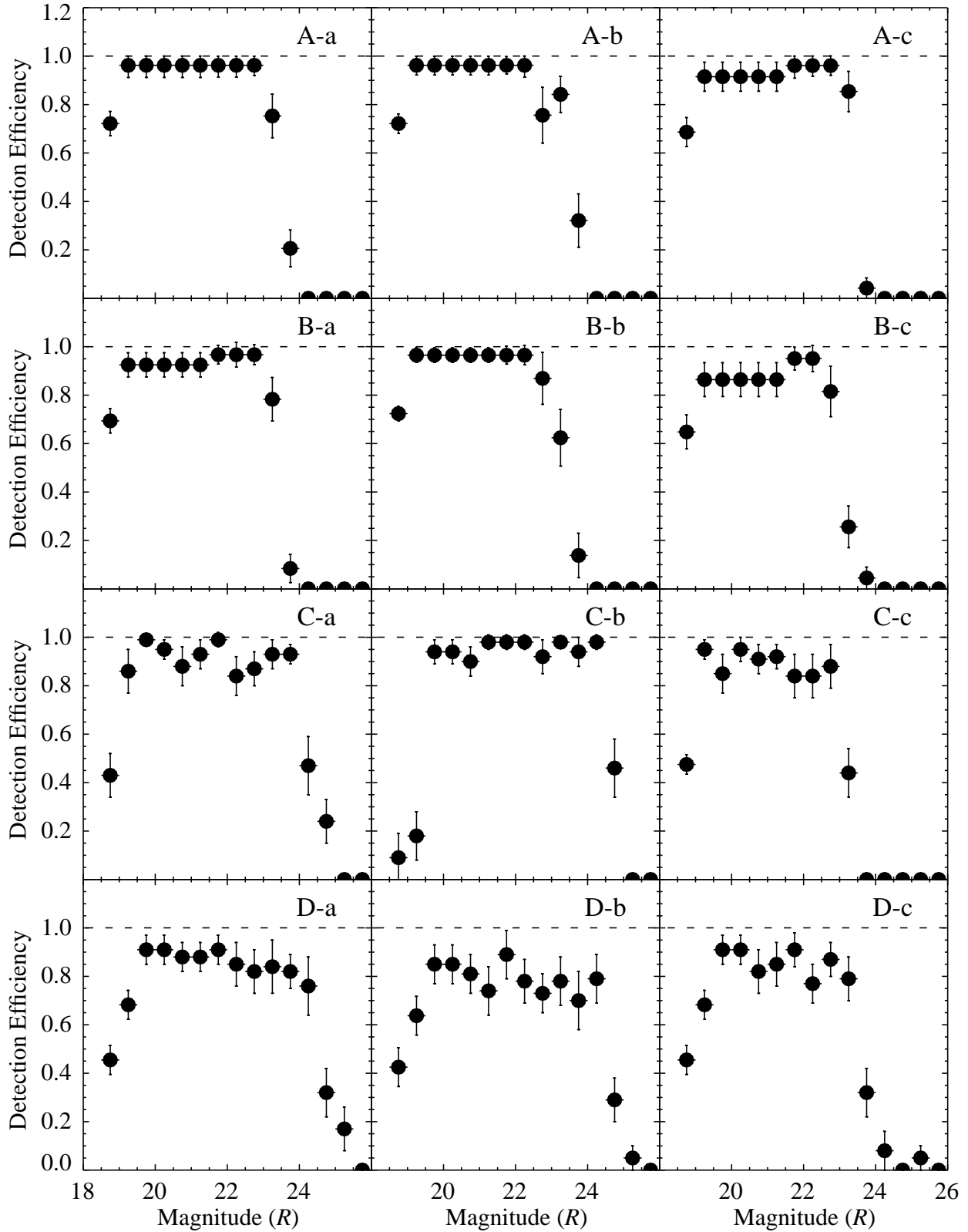}
\caption{
Detection efficiency versus $R$ magnitude of the supernova 
for 12 representative examples 
among the 219 $2\rm{k}\times2\rm{k}$ fields that were searched for SNe.
Supernovae 1995as, 1996cj, 1997ai and 1997ep were respectively discovered on
fields A-a, B-a, C-a and D-a. 
}
\label{figeff}
\end{figure}

We reached significantly fainter detection limits during these
observations compared to the data in Paper~I.  Figure~\ref{figeff}
shows the fractional number of simulated SNe recovered, as a function
of SN detected magnitude, for 12 representative examples among 
the 219 fields observed. For a
typical field the detection efficiency is over 85\% for any stellar
object brighter than $R=23.5$. Note that the loss in efficiency at the
brightest magnitudes is due to detector saturation for bright sources.
The plateau efficiency seen at intermediate magnitudes simply reflects
the areal coverage lost due to masking of the region surrounding bright 
stars.

\begin{figure}
\figurenum{2}
\epsscale{0.9}
\plotone{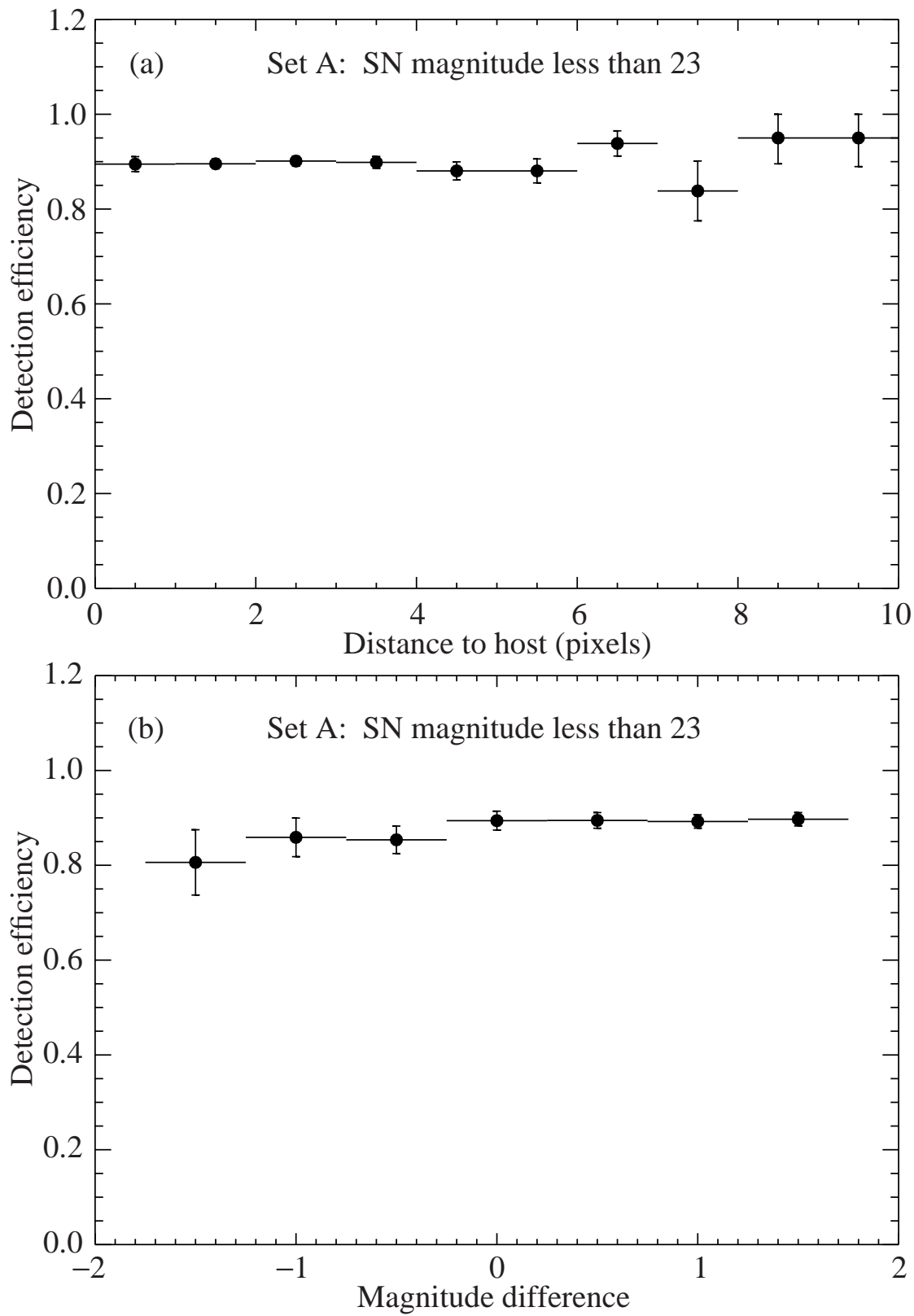}
\caption{(a) detection efficiency versus projected distance to host galaxy;
(b) detection efficiency as a function of 
magnitude difference 
between the host galaxy and the supernova (Host $R$-mag - SN $R$-mag). 
In both plots, an overall $\sim10\%$ inefficiency is present, due to the 
areal coverage lost by masking the region surrounding bright stars, 
independently of the distance to host or the magnitude difference.
}
\label{figeff2}
\end{figure}

The efficiency depends primarily on
the SN magnitude, but the Monte-Carlo simulation also permits to 
account for the small dependence of SN visibility on the host galaxy
surface brightness underlying each SN. This is shown on 
Figure~\ref{figeff2}a where the overall supernova detection efficiency
for Set A is plotted as a function of the magnitude difference between the 
host galaxy aperture flux at the supernova position and the supernova flux.
Figure~\ref{figeff2}-b shows the overall supernova detection 
efficiency as a function of the projected distance to the host galaxy 
center. The detection efficiency does not depend on the SN 
position relative to the host, demonstrating the ability of image subtraction 
techniques to detect supernovae on the nuclei of galaxies.

\paragraph*{Control Time}
We computed a control time as a function of redshift and host galaxy 
magnitude equal to the weighted sum of the number of days during
which the SN could be detected, given the time separation of the 
search and reference images,
where the weighting is according to the corresponding detection efficiency.

Type Ia SN light curves are not unique. The total range for SN~Ia 
$B-$band peak brightness spans $\sim0.5$~mag 
\cite[]{Saha:1999, Gibson:2000}. This has to be taken into account when
computing the control time. Furthermore, as first noted by 
\citet{Phillips:1993}, brighter supernovae also have wider light curves. 
This
correlation between light curve shape and peak luminosity 
has the effect of further increasing the ``visibility'' of 
brighter objects and therefore the time during which they can be detected. 
To account for this correlation, the 
control time was computed, assuming that the SN~Ia light curves form
a one parameter family using an approximation for the 
light curve shape $vs.$ luminosity relation
following the ``stretch factor'' method of \citet[]{Perlmutter:1997a}. 
We assumed that the average SN
light curve follows the average of the best-fit, time-dilated and
$K$-corrected type~Ia template \cite[]{Leibundgut:1988}, with the
generalized cross-filter $K$ correction described by \cite[]{Kim:1996}, 
and that the stretch parameter follows a Gaussian distribution 
with $\sigma\sim0.08$ 
\cite[]{Perlmutter:1999}.
%
The effect of the uncertainties in the light curve shape $vs.$ luminosity
correction and of the remaining $\sim0.15$~mag $B-$band peak luminosity 
intrinsic scatter on the systematic uncertainty in deriving the 
SN rates, is discussed in Section 6.

The SNe\,Ia light curves were
calibrated using Landolt standards \cite[]{Landolt:1992}.  Since
these are observed light-curves, in {\it apparent} magnitudes, no
explicit dependence of our rate on $H_0$, $\Omega_{\rm M}$
or $\Omega_\Lambda$ is introduced at this stage. 
Photometric calibration was not available for all the fields. For those
fields without calibration (about 25\%), zero points were
calculated by comparison with $E$-band (which is close to $R$-band) 
magnitudes of anonymous stars
in the APM (Automated Plate Measuring facility, Cambridge, UK)
POSS-I catalog \cite[]{McMahon:1992}.  
%
A comparison of the APM-$E$ magnitudes with CCD-$R$ magnitudes 
was performed using the fields on
which SNe had been discovered.
The distribution reveals a mean $E-R$ offset of $-0.02$\,mag, with a
dispersion of $0.22$\,mag. Assuming these fields are
representative of the whole dataset, we applied a $0.02$\,mag
shift to the APM magnitudes. The uncertainty in the rate
introduced by this 
uncertain calibration 
is also discussed in Section 6.

Galactic extinction was taken into account for each field
separately using two different methods. Firstly, we used the
Galactic reddening value for each field $E(B-V)$ supplied by D. Burstein
(private communication), derived from the analysis of
\citet[]{Burstein:1982}. We applied these to the data assuming
$R_V=3.1$, and $A_R/A_V=0.751$ \cite[]{Cardelli:1989}. For the
second method, we computed the extinction using more recent
estimations of dust reddening \cite[]{Schlegel:1998} and
$A_R/E(B-V)=2.63$ (appropriate for the Landolt $R$ filter).
Although the individual field reddening values so determined can
sometimes differ by a large amount even 
for our high latitude survey fields, the net effect on the rate is
small as discussed in Section 6.

\section{SN\,Ia rates per unit comoving volume}

To calculate the observed SN\,Ia rate per unit comoving volume, we derive
the expected redshift distribution of SNe, $N_{\rm exp}(z)$, which is
proportional to the observed SN\,Ia rate, $r_{\rm V}(1+z)^{-1}$, where
$r_{\rm V}$ is the rest-frame SN rate per unit comoving volume
and $(1+z)^{-1}$ accounts for cosmological time dilation. The expected
distribution is given by
\begin{equation}
N_{\rm exp}(z)={r_{\rm V}\over{1+z}}\sum_i
S_i \times V(z;{\rm H}_0, \Omega_{\rm M},\Omega_\Lambda)
\times \Delta T_i (z)
\label{eq1}
\end{equation}
where $i$ runs over all the survey fields, $S_i$ is the field
solid angle, and $V(z)$ is the comoving volume element at redshift $z$
(formally $d^2V/dz\,dS$) which depends on the cosmological parameters
$H_0$, $\Omega_{\rm M}$ and $\Omega_\Lambda$ (see, for example,
Equation (26) in \citet[]{Carroll:1992}). Since the supernova
detection efficiency depends on the galaxy apparent magnitudes
$(R_{gal})$, the control time per field at redshift $z$, ($\Delta
T_i(z)$) is computed as $\Delta T_i(z)=\sum_R N_{gal}(z,R)_i\Delta
T_i(z,R_{gal})/ \sum_R N_{gal}(z,R)_i$ where the sum runs over all
possible galaxy apparent magnitudes.
%
Individual control times have been calculated for each field in
bins of $z$ and $R$ (the size of the bins used is 0.5\,mag in $R$
and 0.1 in $z$).

Twenty seven of our 219 search fields had been chosen specifically 
to target high-redshift clusters. 
Suitable clusters and their redshifts were taken from
\cite[]{Gunn:1986}.  
Although clusters will be found quite
naturally in the SCP wide-field images, it is conceivable that
they are over-represented. 
For each cluster target field, we therefore
determined the excess number count as a function of $R$ magnitude
and cluster-centric radius by subtracting the appropriate
background field. Assigning the known redshift of the
appropriate cluster to the excess populations so determined, the
effect on the rate per unit comoving volume was estimated by increasing
the comoving volume element at the cluster redshift by the fractional
excess of luminosity. The uncertainty in the rate introduced 
by targeting these high-redshift clusters is estimated in Section 6.

\paragraph*{One-parameter fits}
Assuming a constant SN\,Ia rate as a function of redshift in the
region covered by these data, we can perform a maximum likelihood 
fit of the observed redshift
distribution to $N_{\rm exp}$ and hence derive $r_{\rm V}$ at a mean
redshift, $\bar{z} = \int{ z N_{\rm exp}(z) dz}/\int{N_{\rm exp}(z)
dz} $.  

The dependence of $r_{\rm V}$ on the Hubble parameter $H_0$ is easily
factorized (since the comoving volume element scales as $H_0^{-3}$), but
$r_{\rm V}$ also depends on the cosmological parameters $\Omega_{\rm
M}$ and $\Omega_\Lambda$.  At $z=0.5$, the comoving volume element in a
flat universe with $\Omega_\Lambda=0.7$ is twice that in a flat
universe with no cosmological constant.  Table~\ref{tabvolfit}
gives the results of the fits for different values of $\Omega_{\rm
M}$ and $\Omega_\Lambda$.  For a spatially-flat cosmological model with
$\Omega_{\rm M}=0.28$ as measured by the SCP 
\cite[]{Perlmutter:1999} and also reported by the High-Z search Team 
from their complete set of spectroscopic SNe~Ia
\cite[]{Riess:1998}, we obtain
\begin{equation}
r_{\rm V}=1.53^{+0.28}_{-0.25}({\rm stat})~10^{-4}~h^3\,{\rm Mpc}^{-3}\,{\rm yr}^{-1}
\label{eq2}
\end{equation}
where the error is statistical only at this stage and $h=H_0/100$.
Slightly different results for $\Omega_{\rm M}$ have also been reported by 
both groups depending on the sample retained in the analysis and the 
method used and 
$\Omega_{\rm M}$ is also been measured with different techniques   
(see for example
\citet[]{Peacock:2001}).
It is therefore interesting to investigate the effect on the rate
of changing the values of the cosmological parameters.
A closer inspection of the comoving volume element dependence on $\Omega_{\rm
  M}$ and $\Omega_\Lambda$ shows that, to a good approximation
($<5$\%), this quantity depends only on the difference
$\omega=\Omega_{\rm M}-\Omega_\Lambda$ in our redshift range and
for $0.1<\Omega_{\rm M}<1.5$ and $|\omega|<1.5$. We therefore also
provide the result as a function of $h$ and $\omega$
and find that the following is a good approximation to our
results:
\begin{equation}
r_{\rm V}(\bar{z}=0.55) =
[(2.06^{+0.37}_{-0.33})\times(1+0.58\omega)]~10^{-4}~h^3\,{\rm Mpc}^{-3}\,{\rm yr}^{-1}
\label{eq3}
\end{equation}
where the error is again only statistical at this stage.

\begin{figure}
\figurenum{3}
\plotone{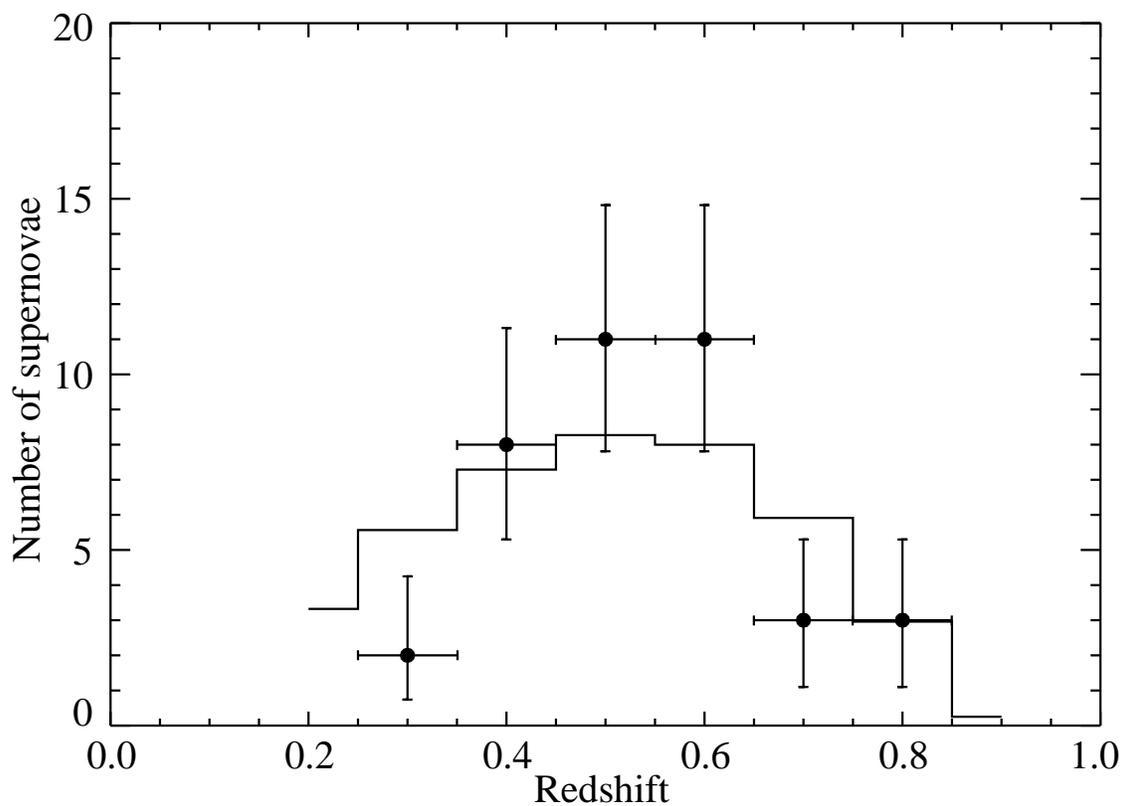}
\caption{
SN~Ia Rate per unit comoving volume: Comparison of Monte-Carlo 
calculation (histogram)
and data (points) for the number of observed
SNe as a function of redshift. A value of
$1.53~h^3~10^{-4}~{\rm Mpc}^{-3}~{\rm yr}^{-1}$
is assumed for the rate.
The prediction assumes no evolution for the rate per unit comoving volume
computed with $\Omega_{\rm M}=0.28$ and $\Omega_\Lambda=0.72$.
}
\label{fignsnobs}
\end{figure}

A comparison
of the expected number of SNe and the observed number is shown in
Figure~\ref{fignsnobs} where the expected number has been computed
assuming no evolution for the rate per unit comoving volume 
and a flat universe
with $\Omega_{\rm M}=0.3$. The agreement is quite good, although the
expected distribution is slightly flatter.

\begin{figure}
\figurenum{4}
\plotone{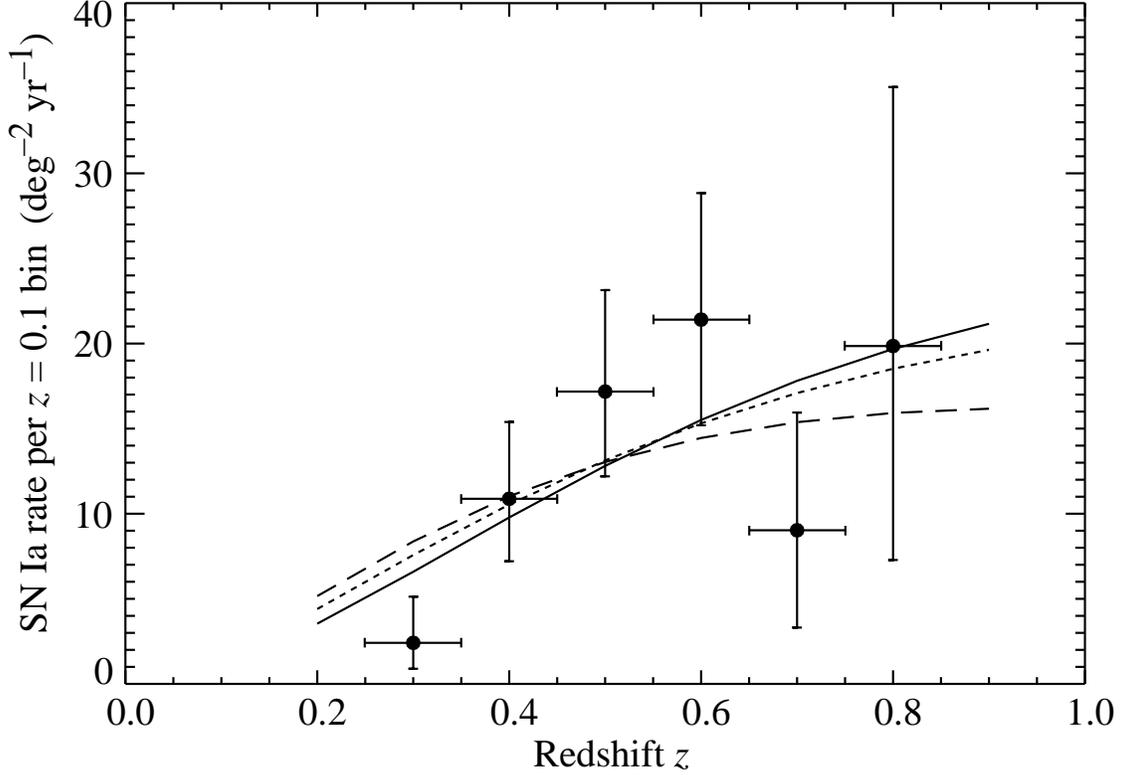}
\caption{
Number of SNe~Ia per z=0.1 redshift bin per square degree per year
as a function of redshift.
Over plotted (lines) are the predictions assuming
that
the number of SNe is proportional to the comoving volume and
adjusted to best fit the observed number of SNe
between $z=0.25$ and $z=0.85$. The solid line is for 
a comoving volume given by a cosmological model with $\Omega_{\rm M}=0.28$
and $\Omega_\Lambda=0.72$ while the dotted line is 
for $\Omega_{\rm M}=0.3$ and 
$\Omega_\Lambda=0.0$ and the dashed line is for $\Omega_{\rm M}=1.0$ and
$\Omega_\Lambda=0.0$.
}
\label{fignsnvol}
\end{figure}

Using the above determination of $r_{\rm V}$, one can compute the
theoretical number of SNe that are produced as a function of
redshift. This is shown in Figure~\ref{fignsnvol} where the number
of SNe per square degree and per year is plotted as well as
predictions for different cosmological models adjusted to best fit the
observations (assuming that the number scales with comoving
volume).

\paragraph*{Two-parameter fits}
In the previous paragraph,
the rest-frame SN rate $r_{\rm V}$ is assumed
constant over the redshift range of interest. Several 
studies have addressed the expected variation of the Ia rate with
redshift (see for example
\citet[]{Madau:1998,Ruiz-Lapuente:1998,Sadat:1998}). With our
enlarged sample spanning the redshift range $0.25$--$0.85$, it is
possible to consider an observational constraint on possible rate
evolution. 
We choose to approximate any potential evolution with a
power law of the form $r_{\rm V}(z)=r_{\bar{z}}[(1+z)/(1+\bar{z})]^\alpha$, 
where $r_{\bar{z}}$ is the
$z=\bar{z}$ type Ia rate per unit comoving volume and $\alpha$ is an index of
evolution ($\alpha=0$ indicates no evolution). Equation~(\ref{eq1}) then
becomes
\begin{equation}
N_{\rm exp}(z)=\frac{r_{\bar{z}}(1+z)^{\alpha-1}}{(1+\bar{z})^\alpha} \sum_i
S_i \times V(z;{\rm H}_0, \Omega_{\rm M},\Omega_\Lambda)
\times \Delta T_i (z)
\label{eq4}
\end{equation}
and we perform a two parameter fit of $r_{\bar{z}}$ and $\alpha$.  

As before, we perform maximum likelihood 
fits for a choice of cosmological models. The results are reported in 
Table~\ref{tabvolfit}.  As expected, the
evolution parameter $\alpha$, depends strongly on the assumed
cosmology. A spatially-flat $\Lambda$ dominated universe 
(model $\Lambda$) favors a solution with little evolution in
the SN\,Ia rate per unit comoving volume, whereas in an Einstein-de Sitter
universe (model E) more evolution is permitted.

\begin{figure}
\figurenum{5}
\plotone{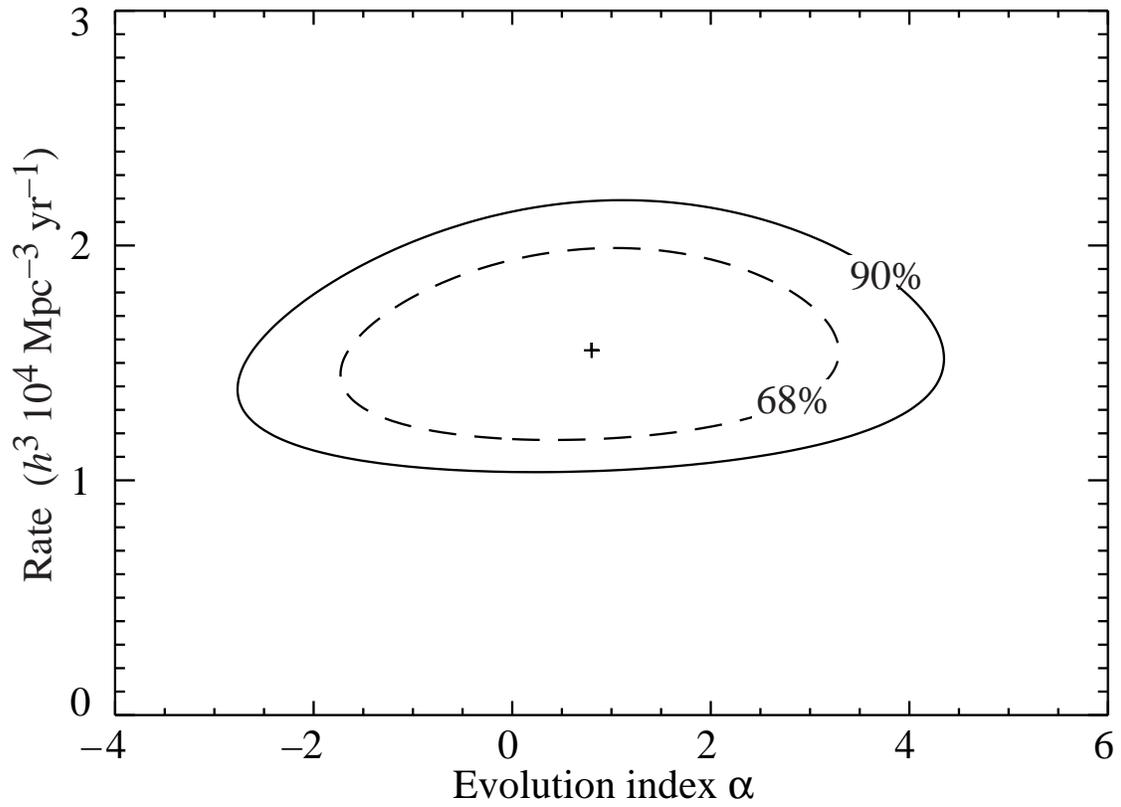}
\caption{Two-parameter maximum likelihood fit of the distant SN~Ia rate:   
68.3\% and 90\% confidence  
regions for the rate per unit comoving volume
versus the evolution index for a comoving volume corresponding 
to a flat universe with $\Omega_{\rm M}=0.28$.
}
\label{figevol}
\end{figure}

For the spatially-flat case (model $\Lambda$) with $\Omega_{\rm
M}=0.28$, we obtain (Fig.~\ref{figevol}) 
\begin{equation}
r_{\bar{z}=0.54}=1.55^{+0.29}_{-0.30}~10^{-4}~h^3~{\rm Mpc}^{-3}~{\rm yr}^{-1}~~~~~{\rm
and}~~~~~\alpha=0.8^{+1.6}_{-1.6} 
\label{eq5}
\end{equation}
where the error is statistical only. 

Although the current dataset does not yet allow a stringent
constraint to be placed on evolution in the SN\,Ia rate, with the
ever-increasing number of SNe found in controlled experiments both
at low and intermediate redshifts, the situation will improve
quite rapidly. The SN\,Ia rate will therefore soon become a key
ingredient in constraining the astrophysical evolution of host
galaxies and in limiting possible progenitor models for SNe\,Ia.

\section{SN Ia rates per unit galaxy luminosity}

Local estimates of the SN\,Ia rate are often expressed in the
``supernova unit'' (SNu), the number of SNe per century per $10^{10}$
solar luminosities in the rest-frame $B$-band. To compare our distant
SN\,Ia rate with any local determinations, one must either convert the
higher-redshift rates into SNu, or convert the local rates into
``events\,\,Mpc$^{-3}$\,\,yr$^{-1}$''. In this section we explore the
former option.

To estimate our rate in SNu, we proceed as described in Paper~I
and calculate the expected redshift distribution of SNe\,Ia given
by
\begin{equation}
N_{\rm exp}(z)={r_{\rm L}\over{1+z}}\sum_i\sum_R N_{gal}(z,R)_i \times L_B
(z,R;{\rm H}_0, \Omega_{\rm M},\Omega_\Lambda) \times \Delta T_i
(z,R)
\label{eq7}
\end{equation}
where $i$ runs over all fields, $R$ is the galaxy apparent
$R$-band magnitude, and $L_B$ is the galaxy rest-frame $B$-band
luminosity in units of $10^{10}$\Lbsun~ which depends upon the
cosmological parameters $H_0$, $\Omega_{\rm M}$ and
$\Omega_\Lambda$.

Since thousands of anonymous high-redshift galaxies are observed
in every survey image, it is more difficult than in local SNe
searches to estimate the number, morphological type and luminosity
distributions of galaxies searched within a given redshift range.
To utilize a determination of the total $B$-band galaxy
luminosity, as a function of $z$ and apparent magnitude $R$, it
will also be necessary to have the relevant galaxy $K$-corrections needed
to convert observed $R$ magnitudes into rest-frame $B$ magnitudes.

We approach this determination of $L_B(z,R)$ in two ways. Firstly,
as in Paper~I, we use observed $R$-band galaxy counts as a
function of redshift and compute from these the rest-frame
$B$-band galaxy luminosity. 
As a second estimate, we compute $L_B(z,R)$ by integrating
recently-determined luminosity functions parameterized via the
Schechter function. We adopt $\Mbsun=5.48$.

\paragraph*{Utilizing CFRS galaxy counts}
$R$-band counts as a function of redshift were kindly calculated
by Simon Lilly \citep{Lilly:1995a} and are based on the analysis
of $I$-band magnitude--redshift data obtained in the Canada-France
Redshift Survey \citep[and references therein]{Lilly:1995b}.
Since the $I$-band is fairly close to the $R$-band, and the
magnitude range of the CFRS sample is comparable to that of our
data,
the extrapolation is small and therefore the dependence of
$R$-band counts on the cosmological parameters is negligible.  To
compute the rest-frame $B$-band galaxy luminosities from apparent
$R$ magnitudes, we used $B-R$ colors and $B$-band $K$-corrections
provided by \citet[]{Gronwall:1995}.

The SN\,Ia rate per unit luminosity was then derived using this
estimate of $L_B(z,R)$ assuming that the rate per unit luminosity
is constant as a function of redshift (an assumption we
investigate in $\S$6). The result is reported in
Table~\ref{tablumfit}.

\paragraph*{Utilizing observed luminosity functions}
We also estimated $L_B(z,R)$ by integrating recently-derived
Schechter parameterizations of local field galaxy luminosity
functions (LFs). We adopted a set of type-dependent LFs covering
three broad galaxy classes; E/S0, spirals, and irregular systems.
Many type-dependent LFs can be found in the literature
\citep{Loveday:1992,Marzke:1998,Folkes:1999}, based on many
different surveys and classification techniques. The agreement is
not particularly good and this renders our calculation somewhat
uncertain (see, for example, \citet{Brinchmann:1999}). Bearing
this in mind, we adopted the LFs of \citet{Marzke:1998} as a
reasonable ``average''.
Type-dependent $K$ and luminosity evolutionary corrections were
adopted from the synthesis models of \citet{Poggianti:1997}.
Finally, to apply these local LFs to higher-redshift samples, we also
need to account for possible evolution in the LFs themselves. The
primary signal is a marked increase with redshift in the abundance
of galaxies with irregular morphology 
which we account for 
by introducing an evolution in the space-density of irregular systems,
adjusted to match the evolution seen by \citet{Brinchmann:1998}.

\begin{figure}
\figurenum{6}
\plotone{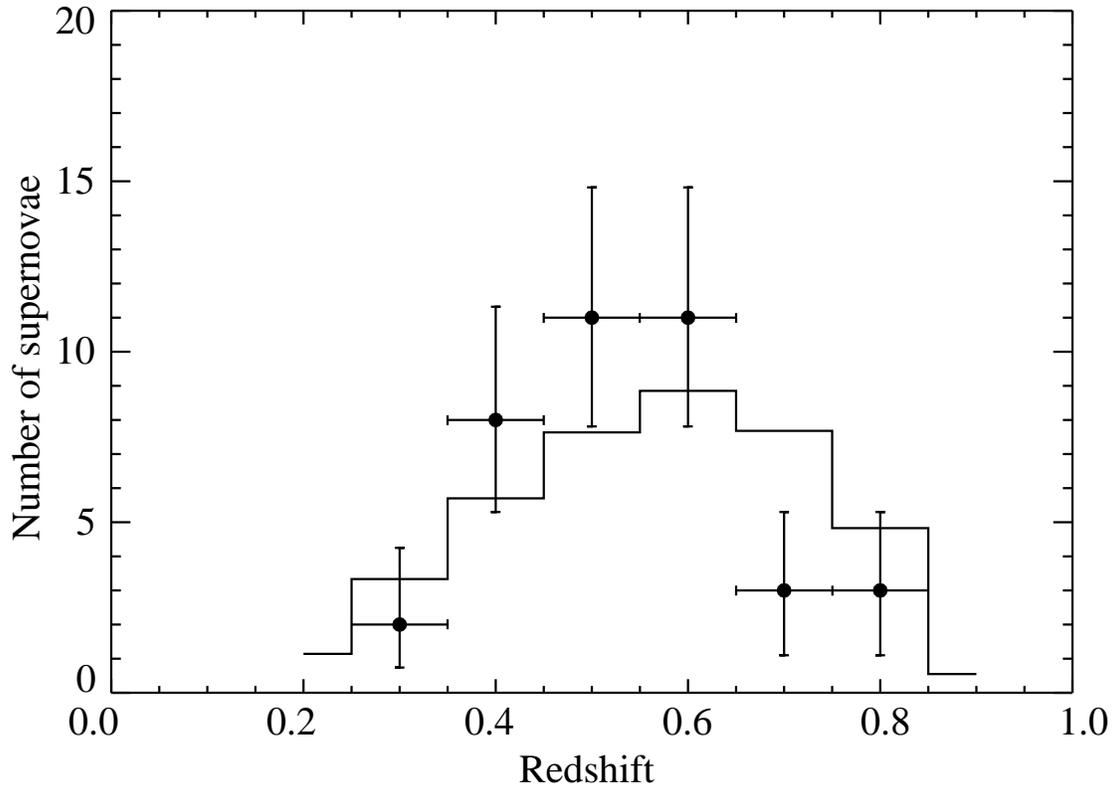}
\caption{
SN~Ia Rate per unit luminosity: 
Comparison of Monte-Carlo calculation (histogram)
and data (points) for the observed number of SNe as a function
of redshift. The prediction assumes that the rate follows the galaxy luminosity
evolution as a function of redshift. A value of $0.58~h^2$~SNu is 
assumed
for the rate, and $\Omega_{\rm M}=0.28$ and $\Omega_\Lambda=0.72$ is used.
}
\label{fignsnlum}
\end{figure}

Figure~\ref{fignsnlum} shows the expected redshift distribution of
SNe\,Ia, $N_{\rm exp}(z)$, as calculated above for a spatially-flat
$\Lambda$ dominated
cosmology (model $\Lambda$) assuming that the SN\,Ia rate per unit
luminosity does not evolve. Similar distributions were computed
for different cosmological models and the rest frame SNe rate
$r_{\rm L}$ was derived by fitting the redshift distribution of
observed SNe to the expected distribution, $N_{\rm exp}(z)$. Results
of these fits are given in Table~\ref{tablumfit}.
For a flat universe with $\Omega_{\rm
M}=0.28$, we find : 
\begin{equation}
r_{\rm L}(\bar{z}=0.55) =
0.58^{+0.10}_{-0.09}({\rm stat})\,h^2\,\rm{SNu} 
\label{eq8}
\end{equation}

The value obtained for model $\Lambda$ is in reasonable agreement (better
than 10\%) with the value obtained from the CFRS galaxy counts.
This is because both estimates of the galaxy luminosity agree very
well in the region $z=0.4$--$0.6$, where most of the SNe were found.
Nevertheless, sizable differences exist in the high redshift
region, where the luminosity derived from the CFRS counts lies
significantly below that derived from the direct LF approach,
probably due to the evolving population of (blue) irregular
systems.  Since a simple extrapolation was used to estimate the
counts at high redshifts from the CFRS data, whereas the
luminosity estimated from the parameterization of LFs used more
recent high redshift survey data, the latter should be more
realistic.
%
\section{Systematic Uncertainties}

With a total of almost forty SNe\,Ia, the statistical uncertainty
is sufficiently small to demand a careful analysis of possible
systematic uncertainties. We estimated these below and summarize
their contribution in Table~\ref{tabsys}.

\paragraph*{Cosmological parameters}
With the methods used in this paper to calculate the SN rates, the
dependence on the cosmological parameters appears only in the
calculation of the comoving volume element or in estimating the absolute
galaxy luminosity. In both cases, the H$_0$ dependence can be
simply factorized.  The dependence on $\Omega_{\rm M}$ and
$\Omega_\Lambda$ is more difficult to derive, although to a very
good approximation ($\simeq5$\%) the comoving volume element depends only on
the combination $\Omega_{\rm M}-\Omega_\Lambda$ in our particular
redshift range (assuming $0.1<\Omega_{\rm M}<1.5$ and
$|\Omega_{\rm M}-\Omega_\Lambda|<1.5$, see $\S$4). In the specific
case of a spatially-flat cosmology, using the SCP value of
$\Omega_{\rm M}=0.28^{+0.10}_{-0.09}$ where statistical and
systematic uncertainties have been combined, the uncertainty on
the event rate becomes
$^{+0.25}_{-0.23}~h^3~10^{-4}~{\rm Mpc}^{-3}~{\rm yr}^{-1}$. For
the rate per unit luminosity (i.e. in SNu), a simple
parameterization on $\Omega_{\rm M}-\Omega_\Lambda$ is not
possible, and for a flat universe with $\Omega_{\rm
  M}=0.28^{+0.10}_{-0.09}$ we find a contribution of
$^{+0.04}_{-0.03}~h^2$~SNu.

\paragraph*{Detection and identification efficiencies}
The study of detection efficiencies as a function of SN magnitude
is an essential element of this analysis. The detection
efficiencies depend upon many parameters and vary widely from
field to field. Uncertainties were determined using a
statistically-limited Monte-Carlo simulation where 250 fake SNe
were added to each image incorporating distribution functions for
the galactocentric distance of each SN and the host galaxy
magnitude distribution (assumed to be representative of the total
galaxy population). The systematic uncertainties were estimated by
varying the 
parameters of the simulation around their nominal
values, and provide a fractional error on the efficiency of less
than 5\% or $\pm 0.08~h^3~10^{-4}~{\rm Mpc}^{-3}~{\rm yr}^{-1}$
($\pm 0.03~h^2$~SNu) for each contribution.

During the SCP SN search, differenced images of candidates
satisfying loose cuts were scanned by eye and candidates kept or
rejected following some quality criteria. This could give some
systematic effects which are {\em a priori} difficult to estimate
precisely.  However for the rate analysis, an automatic procedure
was used to retrieve the few hundred fake SNe that were added to
each field in order to compute the detection efficiency. This
makes it possible to estimate the SCP ``scanning'' efficiency.
Interestingly, we found it to be better than 98\% inside the
nominal cuts which were set higher than during the actual search.

On the other hand, as discussed in Section 3, one SN candidate was not
confirmed spectroscopically (see Table~\ref{tabsn}) and nine others 
were only spectroscopically identified as Type~I and retained as   
``probable Ia'' based on a combination of factors: two have E/S0
hosts and they all have light curve shape and magnitude at peak 
compatible with that of a Type~Ia supernova. Furthermore, since 
no Ib/c was identified in the sets C and D from the 20 candidates
which had spectra, the
probability of having a Type Ib/c supernova in Set A and B, 
where the unidentified candidates have been found, is less 
that 10\% with 90\% confidence level. We therefore conclude that 
at most one of these 10 candidates which could not be positively 
identified as Ia could be a contaminant.  

All together, we estimate the
systematic uncertainty of both ``scanning'' efficiency and
misclassification effects in the current dataset to be 
~$\pm0.03~10^{-4}~h^3~{\rm Mpc}^{-3}~{\rm yr}^{-1}$ or 
~$\pm0.01~h^2$~SNu.
In combination, uncertainties in the detection efficiencies
(detection, scanning, misclassification) translate into a overall
systematic uncertainty on the rates of
~$\pm0.12$ $10^{-4}$~$h^3$~${\rm Mpc}^{-3}$~${\rm yr}^{-1}$ 
~($\pm0.04~h^2$~SNu).

\paragraph{Range of SN\,Ia light curves} 
Control times were calculated following the procedure described in 
Section 3. Noticing that the different 
implementation of the light curve shape $vs.$ luminosity correlation
\cite[]{Hamuy:1996, Phillips:1999, Riess:1996, Riess:1998, Perlmutter:1997a} 
can give somewhat different corrections~\cite[]{Leibundgut:2001},
we conservatively estimated the 
systematic uncertainty coming from the light curve shape
$vs.$ luminosity correlation by varying the ``stretch'' 
parameter by one standard deviation around its nominal value.   
The effect on the rate was found to be
$0.13~h^3~10^{-4}$\,Mpc$^{-3}$\,yr$^{-1}$ ($0.04~h^2$~SNu).  
The ``intrinsic'' scatter of $0.15$\,mag translates into a change in
the rate of $0.03~h^3~10^{-4}$\,Mpc$^{-3}$\,yr$^{-1}$
($0.02~h^2$~SNu). The
overall uncertainty due to the dispersion of SN\,Ia
light curves therefore amounts to
$\pm0.14~h^3~10^{-4}$\,Mpc$^{-3}$\,yr$^{-1}$ ($\pm 0.05~h^2$~SNu)
on the rate.

\paragraph*{Field calibration}
Measured SN light curves, calibrated with Landolt standards, were used
to compute the control time. Efficiency versus magnitude curves, also
needed to compute the control time, were obtained for all fields
calibrated with Landolt standards when available (for 75\% of the
fields) or with the APM catalog for the others.  Errors in the APM
calibration of the fields thus alter the determination of the
efficiency as a function of magnitude and therefore the control time.
This has a sizable effect on the derived SN\,Ia rate since at the
magnitude of most of our SNe, the detection efficiency varies rapidly
with magnitude. We estimated the size of the effect by comparing the
discovery magnitudes of our SNe calibrated using Landolt stars and
calibrated with the APM, assuming that this was representative of our
set of fields. Since only 25\% of our fields lack Landolt calibration,
the overall effect is reduced.  It contributes
$\pm0.06~10^{-4}~h^3~{\rm Mpc}^{-3}~{\rm yr}^{-1}$
( $\pm0.02~h^2$~SNu) to the uncertainty in the rate.

\paragraph*{Galaxy luminosity}
The CFRS galaxy counts are based on data that are well-matched to
our survey in magnitude and redshift range, and only minimal
extrapolation was required to convert from the $I$ to $R$ band.
The associated uncertainty should be small and this is supported
by the calculation based on using the observed LFs as discussed in
Section 5 (Table~\ref{tablumfit}). The difference in the two
calculations serves as our estimate of the systematic uncertainty
here and this amounts to $\pm 0.05~h^2$~SNu.

\paragraph*{Contribution from clusters}
Twenty seven of our 219 search fields had been chosen specifically 
to target high-redshift clusters. 
We followed the procedure described in Section 4 to account for the 
excess number counts that could arise from selecting these fields.
Although the procedure may suffer large statistical and systematic
uncertainties, it only affects a small fraction of the overall search
area. We estimated a 50\% overall uncertainty in estimating the excess 
number counts. This translates into less than a 10\% uncertainty
in calculating the overall contribution to the galaxy counts from
clusters, giving a contribution of $\pm 0.02~h^2$~SNu ($\pm
0.05~10^{-4}~h^3~{\rm Mpc}^{-3}{\rm yr}^{-1}$) to the uncertainty in
the rate. It is likely this is an over-estimate of the uncertainty
given we expect clusters to occur within typical survey fields.

\paragraph*{Galactic extinction}
Galactic extinction was computed using two different methods, one
taken from \citet{Burstein:1982} based on emission from atomic
neutral hydrogen, and the other from \citet{Schlegel:1998}, based
on dust emission in the far infrared (FIR).  Both groups quote 10\%
uncertainty in their estimate of the reddening but differences as
big as a factor of $2$ in $E(B-V)$ were found.  However, the FIR
emission maps have much better resolution which can be important
in our case where the reddening has to be known for specific lines
of sight. We therefore used \citet{Schlegel:1998} as our baseline.
The effect on the rate is nevertheless very small since most of
our fields were selected to have little or no reddening. Overall
the $\pm10$\% uncertainty in the reddening translates into an
uncertainty on the rate of $\pm
0.02~10^{-4}~h^3~{\rm Mpc}^{-3}~{\rm yr}^{-1}$ ($\pm 0.01~h^2$~SNu).

\paragraph*{Host galaxy inclination and extinction} The effect of host
galaxy inclination on our detection efficiency and galaxy
luminosity estimates should be taken into account when calculating
SNe rates. \citet[]{Cappellaro:1999} recently re-estimated the
inclination correction factors for relevant nearby searches.  In
this analysis, both the search technique (in our case subtraction
of CCD images) and calculation of the galaxies' luminosities were
performed in a different manner than in most local searches, so
the inclination effects may not be the same. Inclination and
extinction would reduce both the number of SNe detected {\em and}
the galaxy visible luminosity whose effects may partially cancel
in estimating the rate. A complete analysis of this effect would
require careful modeling of galaxy opacities, which is beyond the
scope of this paper.  Our result should therefore be directly
compared with uncorrected values derived in nearby searches, with
particular attention to CCD searches.

\paragraph*{Brightness evolution and intergalactic dust}
The effect of possible SN~Ia brightness evolution or  
presence of intergalactic dust was not explicitly taken into
account in our derivation of the rates. However, since the SN~Ia 
light curves used to compute the detection efficiencies were 
calibrated using the observed light curves, a possible difference in 
the brightness of distant SNe~Ia compared to local ones is
taken into account whether it is due to evolution or cosmology. 
On the contrary, possible presence of intergalactic dust would
have the effect of lowering the number of observed supernovae. In that
case our results would have to be interpreted as a lower limit of the 
true distant rate. 

\section{Discussion}

We have derived a rest-frame SN\,Ia rate per unit comoving volume at redshift
range $0.25$--$0.85$ ($\bar{z}\simeq0.55$) of
\begin{equation}
r_{\rm V}(\bar{z}=0.55) =
[(2.06^{+0.37}_{-0.33}({\rm stat})\pm0.20({\rm syst}))\times(1+0.58\omega)]~10^{-4}~h^3\,{\rm Mpc}^{-3}\,{\rm yr}^{-1}
\label{eq9}
\end{equation}
with $\omega=\Omega_{\rm M}-\Omega_\Lambda$, where the first
uncertainty is statistical and the second includes systematic effects
which are independent of the systematics arising from the uncertainty
on the cosmological parameters.

For a spatially-flat universe consistent with the SCP results
(i.e. with $\Omega_{\rm M}=0.28^{+0.10}_{-0.09}$ (statistical and
systematic uncertainty combined quadratically)), we measure~: 
\begin{equation}
r_{\rm V}^{\rm flat}(\bar{z}=0.55)=
1.53^{+0.28}_{-0.25}({\rm stat})^{+0.32}_{-0.31}({\rm syst})~10^{-4}~h^3~{\rm Mpc}^{-3}~{\rm yr}^{-1}
\label{eq11}
\end{equation}
 where the systematic uncertainty includes the uncertainty on
the cosmological parameters.

As most low-redshift determinations of the SN\,Ia rate are
reported in `SNu', we also estimate our SNe rate in these units.
For the rate per unit luminosity, we obtain the following result
\begin{equation}
r_{\rm L}(\bar{z}=0.55) =
0.58^{+0.10}_{-0.09}({\rm stat})^{+0.10}_{-0.09}({\rm syst})~h^2\,{\rm SNu}
\label{eq12}
\end{equation}
for a flat universe with $\Omega_{\rm M}=0.28^{+0.10}_{-0.09}$,
and for an Einstein de Sitter universe, we measure
$
r_{\rm L}(\bar{z}=0.55) = 0.94^{+0.16}_{-0.14}({\rm stat})\pm0.14({\rm syst})~h^2\,{\rm SNu}
$
in good agreement with our first measurement reported in Paper~I,
based on the discovery of 3 SNe\,Ia at
$z\simeq0.4$, of
$r_{\rm L}(\bar{z}=0.4)=
0.82^{+0.54}_{-0.37}({\rm stat})^{+0.37}_{-0.25}({\rm syst})~h^2\,{\rm SNu}$

We have studied the redshift dependence of the rate per unit comoving 
volume and
put constraints on the rate of evolution of the SN\,Ia rate. 

\paragraph*{Comparison with other estimates}
In a recent work to be submitted for publication
\cite[]{Reiss:2002}, D.J. Reiss reports values for the SN~Ia rate per unit
luminosity and per unit volume in excellent 
agreement with our values ($<1\sigma$). His values are based on a
sample of 20 SNe at a mean redshift $z\sim0.49$.
Local
$z\simeq0.01$ SN\,Ia rates have been recently reanalyzed, combining
data from five SNe searches (see \citet[]{Cappellaro:1999} and
references therein).  
They find $r_{\rm L}(z=0.01)=
0.36\pm0.11\,h^2\,\rm{SNu}$, 
averaged over all galaxy
types. The quoted uncertainties include estimate of systematic
effects. The SN\,Ia rate at $z\simeq0.1$ has also been measured by
\citet{Hardin:2000}, who find $r_{\rm L}(z=0.1)=
0.44^{+0.35}_{-0.21}\,^{+0.13}_{-0.07}\,h^2\,\rm{SNu}$ (here
systematic and statistical errors are quoted).

In comparing these rates with our measurements, one should bear in
mind the following caveats: (1) most local measurements 
(e.g. in \citet{Cappellaro:1999} have been
based on photographic data rather than CCD data as used here, (2)
we did not apply any correction for host galaxy absorption and
inclination, (3) at high redshift, the mix of galaxy types is
likely to be very different (which will affect comparisons if
different types have differing star-formation histories and hence
SN\,Ia rates), and (4) local SN\,Ia rates are typically reported
in SNu, whereas the high-redshift values are more conveniently
calculated in ``events\,\,Mpc$^{-3}$\,\,yr$^{-1}$'' as the
rest-frame $B$-band luminosity is difficult to estimate.

\begin{figure}
\figurenum{7}
\plotone{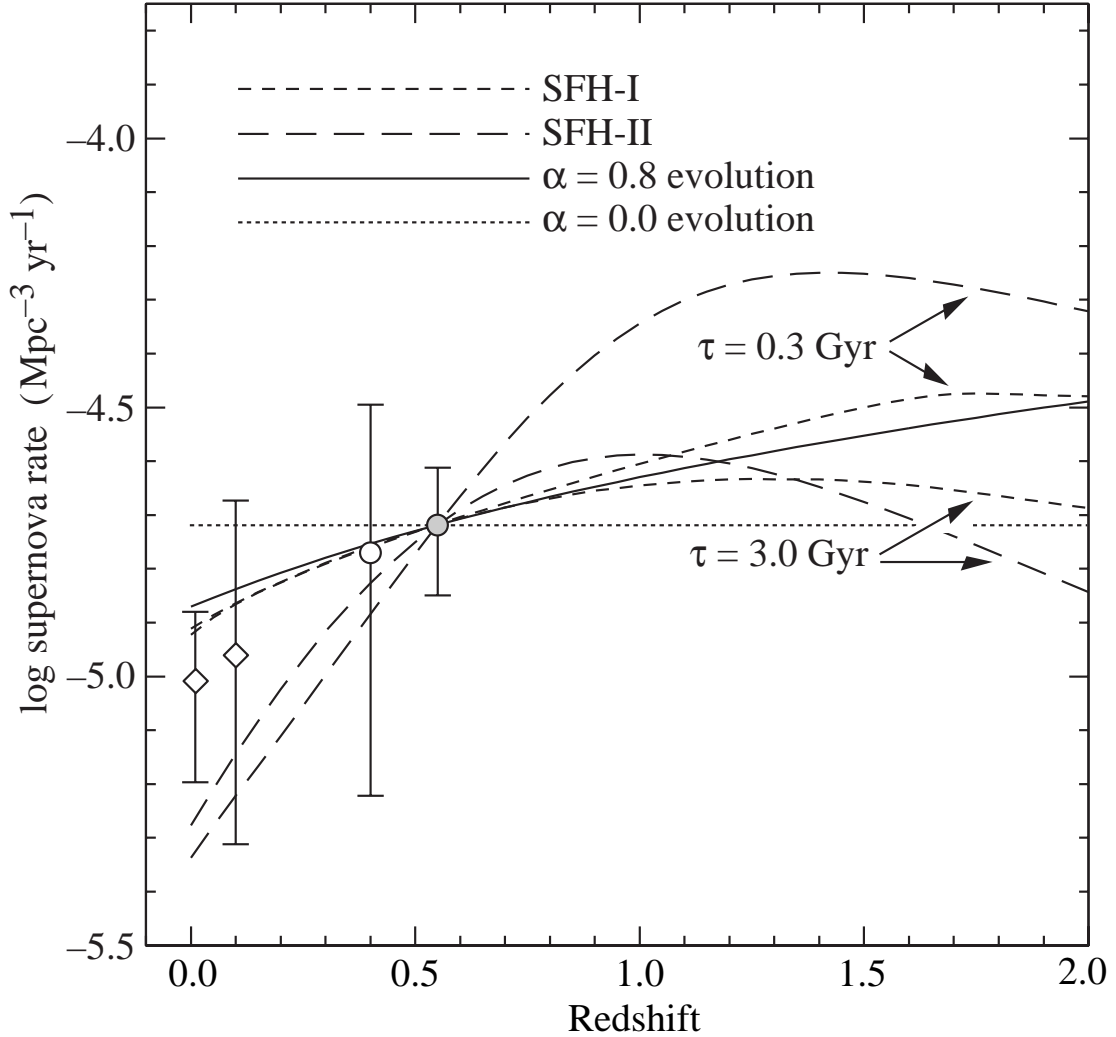}
\caption{
  The SN~Ia rate per comoving volume 
  determined here (filled circle) compared 
  with that of Paper~I (open circle) and of 
  \citet{Cappellaro:1999} at $z\sim0.01$ and \citet{Hardin:2000} 
  at $z\sim0.1$ (open diamonds).
  For comparison, theoretical predictions for two star formation 
  history scenarios and two
  delay times are shown, see text for details. Local SN~Ia rates
  have been converted from SNu units. Also shown are an $\alpha=0.8$
  evolution in the SNe\,Ia rate (solid line) as well as the no evolution
  case (dotted line). The diagram is drawn for
  $H_0=50\,{\rm km}\,{\rm s}^{-1}\,{\rm Mpc}$ and a flat $\Lambda$ 
  dominated model with $\Omega_{\rm M}=0.3$.}
\label{rates_comparison}
\end{figure}

In this section, 
for the purpose of comparing to the models, we convert local rates
from SNu
to ``events\,\,Mpc$^{-3}$\,\,yr$^{-1}$''. To do this, we calculate
the $B$-band luminosity density of the local universe by
integrating local $B$-band luminosity functions
\citep{Marzke:1998,Folkes:1999} and find
$\rho_{L_B}=1.7$--$2.7\times10^8\,h\,L_{B\odot}\,{\rm Mpc}^{-3}$;
we take an average value in this analysis, but note this
introduces a further uncertainty into the calculation. We convert
the local values and plot the results in
Figure~\ref{rates_comparison}.

To this plot we have added recent theoretical predictions for the
form that the evolution of the SN\,Ia rate might take. Various
workers have modeled the expected evolution of the SN\,Ia rate
\citep{Ruiz-Lapuente:1997, Sadat:1998, Madau:1998, Sullivan:2000a}. 
However such work is hampered by
the uncertain physical nature of the progenitor. 
The evolution
expected depends critically on whether SNe\,Ia occur in double or
single degenerate progenitor systems (for a review see
\citet{Nomoto:1999} and references therein), the expected
evolution in the fraction of stellar binaries and, of course, the
cosmic star formation history (SFH).

Here we adopt an empirical approach representing these
uncertainties in terms of two parameters
\citep{Madau:1998,Dahlen:1999}. The first is a delay time $\tau$,
between the binary system formation and SNe explosion epochs,
which defines a (time-independent) explosion probability per white
dwarf.  
Note that this
parameter is treated in different ways in the literature.
\citet[]{Madau:1998} define a (time-independent)
explosion probability per white dwarf, which they model as an exponential
probability function with a mean value of $\tau$ whereas 
\citet[]{Dahlen:1999} use $\tau$ as an \textit{exact} delay time
between binary system formation and supernovae explosion. 
The difference
between the two approaches becomes sizable at higher redshift for scenarios
involving large values of $\tau$. 
Here we adopt the former approach but note that this 
may introduce further uncertainties at $z>1$
in scenarios with large $\tau$.
The second parameter is an explosion efficiency, $\eta$, which
accounts for the fraction of binary systems that never result in a
SNe. We constrain $\eta$ by requiring that our
predicted rate at $z=0.55$ is equal to our new observational
determination. In Figure 6, we show two illustrative values:
$\tau=0.3\,\rm{Gyr}$, corresponding to a shallower decline at
high-redshift, and $\tau=3.0\,\rm{Gyr}$ which produces a steeper
drop-off.

We consider each of these two SNe\,Ia models in the context of two
different star formation history scenarios. The first (SFH-I) is taken 
from
\cite{Madau:2000} who provides a convenient analytical fit of the
star formation rate ($SFR$) form

\begin{equation}
SFR(z)={1.67\times0.23\,e^{3.4z}\over e^{3.8z}+44.7}~M_{\odot}\,{\rm yr}^{-1}\,{\rm Mpc}^{-3}
\label{13}
\end{equation}

\noindent 
in an Einstein-de Sitter universe. We converted this formula to that
appropriate for a $\Lambda$-dominated flat universe by computing
the difference in luminosity density. The SFH fit matches most
$UV$-continuum and H$\alpha$ luminosity densities from $z=0$ to
$z=4$ and includes a mild correction for dust of $A_{1500}=1.2$
mag ($A_{2800}=0.55$ mag). However, the $SFR$
evolution in this
model to $z\simeq1.5$ is both stronger, and results in a lower
local $SFR$, than some recent $UV$ measurements
\citep{Cowie:1999,Sullivan:2000b}. Accordingly, we also consider a
second SFH (SFH-II) with a shallower evolution (a factor of
$\sim4$ from $z=0$ to $z\simeq1.75$ in an Einstein-de Sitter universe, and
constant thereafter).

These various predictions are plotted, for a flat $\Omega_{\rm M}=0.3$ 
cosmology, in Figure~\ref{rates_comparison}, together with our estimate of the
evolutionary index. Although our internal estimate is highly
uncertain, already it would seem to favor scenarios which involve
little evolution over the redshift range $z=0$--$0.6$; a result
which is in agreement with comparisons based on the low redshift
rate determinations. Clearly, a precise measurement of the SN~Ia
rate at (say) $z=1$ would enable further, more robust, constraints
to be placed on any evolution, as the redshift range that is
currently probed is quite small. In the near future, our 
Supernova Cosmology Project's ongoing high redshift supernova searches, 
and those of the High-Z Search Team, 
should provide enough data at these redshifts
to place more stringent constraints on the star formation history.

\section*{Acknowledgments}
The observations described in this paper
were primarily obtained as visiting/guest astronomers at
the Cerro Tololo
Inter-American Observatory 4-meter telescope, operated by the
National Optical Astronomy Observatory under contract to the National
Science Foundation; the Keck I and II 10-m telescopes of  the California
Association for Research in Astronomy;
the Wisconsin-Indiana-Yale-NOAO (WIYN) telescope;
the European Southern Observatory 3.6-meter telescope;
the Isaac Newton and William Herschel Telescopes, 
operated on the island of La Palma by the Isaac
Newton Group in the Spanish Observatorio del Roque de los Muchachos of the
Instituto de Astrofisica de Canarias; the Nordic Optical 2.5-meter
telescope and the   
%
NASA/ESA Hubble Space Telescope 
which is operated by the
Association of
Universities for Research in Astronomy, Inc. under NASA contract No.
NAS5-26555. 
We thank the dedicated staff of these observatories for
their excellent assistance in pursuit of this project.
We thank Gary Bernstein and Tony Tyson for developing and
supporting the Big Throughput Camera at the CTIO 4-meter; this wide-field
camera was important in the discovery of many of the high-redshift supernovae.
We thank Simon Lilly and Caryl Gronwall \& David Koo for providing
their galaxy counts and acknowledge useful discussions with Wal
Sargent, Bruno Leibundgut and Piero Madau.  
This work was supported in part by the
Physics Division, E.~O. Lawrence Berkeley National Laboratory of the
U.~S. Department of Energy under Contract No. DE-AC03-76SF000098, and
by the National Science Foundation's Center for Particle Astrophysics,
University of California, Berkeley under grant No. ADT-88909616.
M.~S acknowledges support from a Particle Physics and Astronomy Research
Council (UK) Fellowship. A.~G. is a Royal Swedish Academy
Research Fellow supported by a
grant from the Knut and Alice Wallenberg Foundation.
The France-Berkeley Fund provided additional collaboration
support.
\newpage
\begin{deluxetable}{lrrcc}
\scriptsize
\tiny
\tablewidth{0pt}
\tablenum{1a}
\tablecaption{Data Set A: fields A-1 to A-23}
\label{tab1a}
\tablehead{
\colhead{Name} &
\colhead{RA(2000)} &
\colhead{~DEC(2000)} &
\colhead{Detection} &
\colhead{E(B-V)} \\
\colhead{~} & \colhead{~} & \colhead{~} 
& \colhead{Limit\tablenotemark{a}} & \colhead{~} 
}
\startdata
A-1 &   1 04 18.51 &   7 46 03.9 &  22.1 &  0.025 \cr
A-2 &   3 07 51.40 &  10 39 42.9 &  22.7 &  0.099 \cr
A-3 &   3 36 59.00 &   0 25 12.7 &  22.8 &  0.114 \cr
A-4 &   3 15 48.75 &  $-1$ 34 39.4 &  23.1 &  0.031 \cr
A-5 &   3 42 32.38 &  17 30 38.9 &  22.5 &  0.128 \cr
A-6 &   1 56 55.65 &   7 42 58.2 &  23.1 &  0.040 \cr
A-7 &   1 55 00.61 &   7 53 37.1 &  23.0 &  0.046 \cr
A-8 &   1 56 28.73 &   8 07 11.5 &  23.1 &  0.043 \cr
A-9 &   1 54 47.22 &   7 55 15.0 &  23.0 &  0.041 \cr
A-10 &   1 54 38.32 &   7 59 21.6 &  23.2 &  0.044 \cr
A-11 &   1 54 28.74 &   8 17 10.7 &  23.2 &  0.044 \cr
A-12 &   1 55 45.31 &   8 15 19.3 &  23.1 &  0.038 \cr
A-13 &   1 53 21.74 &   7 33 23.9 &  23.0 &  0.033 \cr
A-14 &   1 53 32.34 &   7 57 34.0 &  22.9 &  0.038 \cr
A-15 &   1 53 34.85 &   8 18 02.8 &  23.0 &  0.052 \cr
A-16 &   1 33 57.62 &   6 20 25.0 &  23.5 &  0.044 \cr
A-17 &   1 43 10.43 &   2 32 17.6 &  23.0 &  0.023 \cr
A-18 &   1 49 38.77 &   2 04 49.0 &  22.9 &  0.048 \cr
A-19 &   2 03 05.76 &   1 55 26.9 &  23.0 &  0.029 \cr
A-20 &   2 06 59.11 &   6 52 13.1 &  22.7 &  0.035 \cr
A-21 &   2 37 45.13 &   3 43 23.6 &  22.6 &  0.035 \cr
A-22 &   4 59 02.18 &   7 57 55.9 &  23.2 &  0.235 \cr
A-23 &   5 16 46.29 &  14 48 35.0 &  23.4 &  0.688 \cr
\enddata
\tablenotetext{a}{
Defined as the magnitude above which the supernova 
detection efficiency drops below 50\% of the maximum detection efficiency
in the field.
}
\tablecomments{
Col 1: Field Name,
Col 2: Right Ascension (equinox 2000),
Col 3: Declination (equinox 2000),
Col 4: Supernova detection limit, 
Col 5: Color excess from \citet{Schlegel:1998}.
}
\end{deluxetable}

\begin{deluxetable}{lrrcc}
\scriptsize
\tiny
\tablewidth{0pt}
\tablenum{1a}
\tablecaption{Data Set A: fields A-24 to A-46}
\tablehead{
\colhead{Name} &
\colhead{RA(2000)} &
\colhead{~DEC(2000)} &
\colhead{Detection} &
\colhead{E(B-V)} \\
\colhead{~} & \colhead{~} & \colhead{~} 
& \colhead{Limit\tablenotemark{a}} & \colhead{~} 
}
\startdata
A-24 &   8 55 10.82 &   8 01 16.7 &  23.2 &  0.018 \cr
A-25 &   8 26 59.23 &   4 35 37.6 &  23.9 &  0.027 \cr
A-26 &   8 52 05.41 &   2 15 22.2 &  21.8 &  0.034 \cr
A-27 &   2 08 11.00 & $-$13 29 16.8 &  23.2 &  0.019 \cr
A-28 &   1 35 40.23 &   4 23 32.5 &  23.0 &  0.023 \cr
A-29 &   1 37 05.37 &   4 17 45.2 &  23.2 &  0.022 \cr
A-30 &   1 37 37.86 &   4 19 14.0 &  23.1 &  0.021 \cr
A-31 &   1 38 50.55 &   4 21 07.5 &  23.2 &  0.019 \cr
A-32 &   1 39 59.01 &   4 21 19.0 &  23.1 &  0.021 \cr
A-33 &   1 40 51.70 &   4 21 45.3 &  23.2 &  0.022 \cr
A-34 &   1 35 43.81 &   4 30 38.4 &  22.9 &  0.025 \cr
A-35 &   1 36 28.77 &   4 33 17.2 &  22.9 &  0.021 \cr
A-36 &   1 37 24.50 &   4 36 01.0 &  23.1 &  0.021 \cr
A-37 &   1 38 41.33 &   4 27 02.2 &  22.4 &  0.019 \cr
A-38 &   1 39 16.44 &   4 37 20.6 &  23.2 &  0.022 \cr
A-39 &   1 40 44.93 &   4 32 54.6 &  23.2 &  0.019 \cr
A-40 &   3 00 56.51 &   0 28 36.3 &  22.9 &  0.041 \cr
A-41 &   3 03 00.71 &   0 36 25.2 &  22.7 &  0.031 \cr
A-42 &   3 03 58.40 &   0 29 33.5 &  23.1 &  0.029 \cr
A-43 &   3 00 27.57 &   0 52 40.2 &  23.1 &  0.041 \cr
A-44 &   3 01 59.24 &   0 51 06.8 &  23.0 &  0.033 \cr
A-45 &   3 02 36.56 &   0 49 50.4 &  22.8 &  0.033 \cr
A-46 &   3 01 43.69 &   1 01 21.7 &  22.9 &  0.036 \cr
\enddata
\tablenotetext{a}{
Defined as the magnitude above which the supernova 
detection efficiency drops below 50\% of the maximum detection efficiency
in the field.
}
\tablecomments{
Col 1: Field Name,
Col 2: Right Ascension (equinox 2000),
Col 3: Declination (equinox 2000),
Col 4: Supernova detection limit, 
Col 5: Color excess from \citet{Schlegel:1998}.
}
\end{deluxetable}

\begin{deluxetable}{lrrcc}
\scriptsize
\tiny
\tablewidth{0pt}
\tablenum{1a}
\tablecaption{Data Set A : fields A-47 to A-69}
\tablehead{
\colhead{Name} &
\colhead{RA(2000)} &
\colhead{~DEC(2000)} &
\colhead{Detection} &
\colhead{E(B-V)} \\
\colhead{~} & \colhead{~} & \colhead{~} 
& \colhead{Limit\tablenotemark{a}} & \colhead{~} 
}
\startdata
A-47 &   3 22 18.03 &  $-$4 58 15.8 &  23.1 &  0.032 \cr
A-48 &   3 23 13.26 &  $-$4 58 00.5 &  23.1 &  0.034 \cr
A-49 &   5 12 33.79 &  $-$5 28 24.2 &  23.5 &  0.094 \cr
A-50 &   5 14 08.04 &  $-$5 25 26.9 &  23.0 &  0.146 \cr
A-51 &   5 15 06.22 &  $-$5 22 47.4 &  23.5 &  0.142 \cr
A-52 &   5 15 42.22 &  $-$5 27 40.8 &  23.3 &  0.174 \cr
A-53 &   5 16 38.77 &  $-$5 19 49.7 &  23.5 &  0.184 \cr
A-54 &   5 17 29.27 &  $-$5 23 23.6 &  23.3 &  0.191 \cr
A-55 &   5 18 32.34 &  $-$5 27 45.2 &  23.1 &  0.199 \cr
A-56 &   5 11 55.21 &  $-$5 08 31.1 &  23.5 &  0.087 \cr
A-57 &   5 13 13.89 &  $-$5 15 08.8 &  23.5 &  0.125 \cr
A-58 &   5 14 28.75 &  $-$5 15 52.9 &  23.4 &  0.155 \cr
A-59 &   5 15 56.36 &  $-$5 09 13.4 &  23.5 &  0.132 \cr
A-60 &   5 16 21.30 &  $-$5 07 41.7 &  23.5 &  0.120 \cr
A-61 &   5 15 19.46 &  $-$4 52 38.5 &  23.4 &  0.123 \cr
A-62 &   5 15 26.23 &  $-$4 58 06.4 &  23.2 &  0.117 \cr
A-63 &   5 16 54.53 &  $-$4 55 53.2 &  23.5 &  0.091 \cr
A-64 &   8 13 58.35 &  10 02 08.8 &  23.2 &  0.038 \cr
A-65 &   8 16 03.00 &  10 02 51.0 &  23.3 &  0.040 \cr
A-66 &   8 17 32.60 &  10 07 47.0 &  23.0 &  0.035 \cr
A-67 &   8 14 56.29 &  10 11 05.5 &  22.7 &  0.042 \cr
A-68 &   8 15 42.78 &  10 22 30.0 &  23.2 &  0.039 \cr
A-69 &   8 16 58.45 &  10 45 50.5 &  23.2 &  0.037 \cr
\enddata
\tablenotetext{a}{
Defined as the magnitude above which the supernova 
detection efficiency drops below 50\% of the maximum detection efficiency
in the field.
}
\tablecomments{
Col 1: Field Name,
Col 2: Right Ascension (equinox 2000),
Col 3: Declination (equinox 2000),
Col 4: Supernova detection limit, 
Col 5: Color excess from \citet{Schlegel:1998}.
}
\end{deluxetable}

\begin{deluxetable}{lrrcc}
\scriptsize
\tiny
\tablewidth{0pt}
\tablenum{1b}
\tablecaption{Data Set B: fields B-1 to B-23}
\label{tab1b}
\tablehead{
\colhead{Name} &
\colhead{RA(2000)} &
\colhead{~DEC(2000)} &
\colhead{Detection} &
\colhead{E(B-V)} \\
\colhead{~} & \colhead{~} & \colhead{~} 
& \colhead{Limit\tablenotemark{a}} & \colhead{~} 
}
\startdata
B-1 &  12 40 43.23 &  $-$7 09 48.5 &  23.4 &  0.038 \cr
B-2 &  12 34 43.21 &  $-$9 24 52.4 &  23.3 &  0.036 \cr
B-3 &  11 21 33.31 &   0 07 09.2 &  23.5 &  0.043 \cr
B-4 &  10 40 17.51 &  $-$6 59 30.4 &  22.8 &  0.049 \cr
B-5 &  08 54 58.96 &   8 09 17.1 &  23.0 &  0.017 \cr
B-6 &  10 16 42.40 &  $-$1 10 36.9 &  22.4 &  0.031 \cr
B-7 &  08 51 34.90 &   2 16 35.6 &  22.9 &  0.032 \cr
B-8 &  09 00 20.78 &   3 53 52.6 &  22.6 &  0.036 \cr
B-9 &  12 26 48.93 &  11 16 46.7 &  22.6 &  0.033 \cr
B-10 &  12 57 57.65 &  $-$0 38 19.8 &  23.2 &  0.029 \cr
B-11 &  11 32 24.04 &  $-$3 07 30.4 &  23.3 &  0.035 \cr
B-12 &  13 17 29.17 &  $-$4 16 06.3 &  22.3 &  0.024 \cr
B-13 &  14 18 44.14 &   2 52 33.1 &  23.1 &  0.027 \cr
B-14 &  14 19 32.93 &   2 59 42.6 &  23.1 &  0.026 \cr
B-15 &  14 21 00.04 &   2 53 38.3 &  23.1 &  0.024 \cr
B-16 &  14 21 19.88 &   2 55 10.4 &  23.1 &  0.026 \cr
B-17 &  14 22 58.40 &   2 58 46.6 &  23.0 &  0.027 \cr
B-18 &  14 23 54.41 &   2 57 59.6 &  23.1 &  0.026 \cr
B-19 &  14 24 06.19 &   2 57 29.6 &  23.4 &  0.026 \cr
B-20 &  15 04 35.18 &   2 55 42.4 &  23.3 &  0.028 \cr
B-21 &  15 05 51.60 &   2 53 51.3 &  23.4 &  0.030 \cr
B-22 &  15 06 13.93 &   2 56 24.9 &  23.3 &  0.029 \cr
B-23 &  09 56 32.84 &   3 16 54.2 &  22.6 &  0.037 \cr
\enddata
\tablenotetext{a}{
Defined as the magnitude above which the supernova 
detection efficiency drops below 50\% of the maximum detection efficiency
in the field.
}
\tablecomments{
Col 1: Field Name,
Col 2: Right Ascension (equinox 2000),
Col 3: Declination (equinox 2000),
Col 4: Supernova detection limit, 
Col 5: Color excess from \citet{Schlegel:1998}.
}
\end{deluxetable}

\begin{deluxetable}{lrrcc}
\scriptsize
\tiny
\tablewidth{0pt}
\tablenum{1b}
\tablecaption{Data Set B: fields B-24 to B-46}
\tablehead{
\colhead{Name} &
\colhead{RA(2000)} &
\colhead{~DEC(2000)} &
\colhead{Detection} &
\colhead{E(B-V)} \\
\colhead{~} & \colhead{~} & \colhead{~} 
& \colhead{Limit\tablenotemark{a}} & \colhead{~} 
}
\startdata
B-24 &   9 57 24.31 &   3 20 11.9 &  22.2 &  0.042 \cr
B-25 &   9 58 19.95 &   3 20 54.6 &  22.8 &  0.043 \cr
B-26 &   9 56 44.60 &   3 08 34.8 &  22.8 &  0.038 \cr
B-27 &  10 31 46.59 &   0 06 42.0 &  22.6 &  0.038 \cr
B-28 &  10 30 51.46 &   $-$0 06 44.0 &  22.5 &  0.041 \cr
B-29 &  11 23 37.67 &   0 47 12.5 &  23.4 &  0.041 \cr
B-30 &  11 24 39.81 &   0 43 27.8 &  23.1 &  0.040 \cr
B-31 &  13 17 50.56 &   $-$0 09 31.9 &  23.0 &  0.021 \cr
B-32 &  13 19 39.82 &   $-$0 06 45.5 &  22.6 &  0.024 \cr
B-33 &  13 19 59.64 &   $-$0 07 03.1 &  23.2 &  0.025 \cr
B-34 &  13 21 22.75 &   $-$0 08 11.1 &  23.4 &  0.025 \cr
B-35 &  13 22 20.39 &   $-$0 07 01.9 &  23.6 &  0.027 \cr
B-36 &  13 23 04.56 &   $-$0 07 10.9 &  23.5 &  0.033 \cr
B-37 &  13 24 26.99 &   $-$0 06 39.7 &  23.6 &  0.034 \cr
B-38 &  16 06 06.38 &   6 40 14.7 &  22.9 &  0.045 \cr
B-39 &  16 05 59.89 &   6 23 30.0 &  23.3 &  0.049 \cr
B-40 &  16 07 16.97 &   6 26 15.2 &  22.9 &  0.046 \cr
B-41 &  16 08 38.64 &   6 29 30.6 &  23.0 &  0.052 \cr
B-42 &  16 09 07.04 &   6 22 04.6 &  23.2 &  0.051 \cr
B-43 &  16 09 43.78 &   6 26 41.9 &  22.9 &  0.051 \cr
B-44 &  16 10 22.41 &   6 01 20.1 &  23.0 &  0.048 \cr
B-45 &  16 10 47.22 &   5 58 39.0 &  23.4 &  0.050 \cr
B-46 &  16 11 59.78 &   6 00 36.4 &  23.1 &  0.060 \cr
\enddata
\tablenotetext{a}{
Defined as the magnitude above which the supernova 
detection efficiency drops below 50\% of the maximum detection efficiency
in the field.
}
\tablecomments{
Col 1: Field Name,
Col 2: Right Ascension (equinox 2000),
Col 3: Declination (equinox 2000),
Col 4: Supernova detection limit, 
Col 5: Color excess from \citet{Schlegel:1998}.
}
\end{deluxetable}

\begin{deluxetable}{lrrcc}
\scriptsize
\tiny
\tablewidth{0pt}
\tablenum{1c}
\tablecaption{Data Set C}
\label{tab1c}
\tablehead{
\colhead{Name} &
\colhead{RA(2000)} &
\colhead{~DEC(2000)} &
\colhead{Detection} &
\colhead{E(B-V)} \\
\colhead{~} & \colhead{~} & \colhead{~} 
& \colhead{Limit\tablenotemark{a}} & \colhead{~} 
}
\startdata
C-1\tablenotemark{b} &   8 15 49.75 &  10 00 22.4 &  23.8 &  0.040 \cr
C-5 &   8 56 15.99 &   4 41 47.7 &  23.1 &  0.021 \cr
C-9 &   8 59 04.49 &   4 39 53.8 &  23.4 &  0.027 \cr
C-13 &   8 58 34.19 &   4 00 32.8 &  22.9 &  0.029 \cr
C-17 &  11 23 28.58 &   0 56 39.2 &  24.2 &  0.043 \cr
C-21 &  11 31 30.08 &  $-$2 45 35.0 &  24.0 &  0.041 \cr
C-25 &  11 33 28.88 &  $-$2 42 35.2 &  24.0 &  0.037 \cr
C-29 &  11 31 22.77 &  $-$3 17 59.8 &  24.1 &  0.034 \cr
C-33 &  13 20 22.31 &   0 01 09.1 &  24.1 &  0.025 \cr
C-37 &  13 22 37.16 &   0 03 11.9 &  24.5 &  0.026 \cr
C-41 &  14 22 02.21 &   2 51 51.7 &  24.0 &  0.026 \cr
C-45 &  14 24 46.62 &   2 55 49.2 &  23.9 &  0.024 \cr
C-49 &   8 29 48.58 &   5 00 52.2 &  19.6 &  0.019 \cr
C-53 &  10 32 16.36 &  $-$0 12 47.3 &  23.2 &  0.045 \cr
C-57 &  10 35 10.67 &   0 27 23.7 &  23.3 &  0.031 \cr
\enddata
\tablenotetext{a}{
Defined as the magnitude above which the supernova 
detection efficiency drops below 50\% of the maximum detection efficiency
in the field.}
\tablenotetext{b}{
First CCD of the 4 2k$\times$2k Big Throughput Camera.}
\tablecomments{
Col 1: Field Name,
Col 2: Right Ascension (equinox 2000),
Col 3: Declination (equinox 2000),
Col 4: Supernova detection limit, 
Col 5: Color excess from \citet{Schlegel:1998}.
}
\end{deluxetable}

\begin{deluxetable}{lrrcc}
\scriptsize
\tiny
\tablewidth{0pt}
\tablenum{1d}
\tablecaption{Data Set D}
\label{tab1d}
\tablehead{
\colhead{Name} &
\colhead{RA(2000)} &
\colhead{~DEC(2000)} &
\colhead{Detection} &
\colhead{E(B-V)} \\
\colhead{~} & \colhead{~} & \colhead{~} 
& \colhead{Limit\tablenotemark{a}} & \colhead{~} 
}
\startdata
D-1\tablenotemark{b} &   8 58 47.18 &   4 27 12.9 &  24.3 &  0.026 \cr
D-5 &   9 01 26.27 &   4 27 37.9 &  24.5 &  0.027 \cr
D-9 &   9 01 41.03 &   3 49 21.3 &  24.6 &  0.048 \cr
D-13 &   5 37 35.18 &  $-$2 53 03.5 &  22.7 &  0.048 \cr
D-17 &   5 37 33.80 &  $-$3 30 45.1 &  18.6 &  0.065 \cr
D-21 &   5 35 40.87 &  $-$2 26 18.1 &  24.0 &  0.058 \cr
D-25 &   5 35 37.78 &  $-$2 57 02.2 &  24.1 &  0.051 \cr
D-29 &   5 34 46.79 &  $-$3 27 48.5 &  24.4 &  0.045 \cr
D-33 &   5 33 31.54 &  $-$2 14 40.9 &  24.1 &  0.056 \cr
D-37 &   8 59 20.58 &   3 55 52.7 &  24.2 &  0.031 \cr
D-41 &   8 57 00.27 &   4 01 24.3 &  23.6 &  0.026 \cr
\enddata
\tablenotetext{a}{
Defined as the magnitude above which the supernova 
detection efficiency drops below 50\% of the maximum detection efficiency
in the field.}
\tablenotetext{b}{
First CCD of the 4 2k$\times$2k Big Throughput Camera.}
\tablecomments{
Col 1: Field Name,
Col 2: Right Ascension (equinox 2000),
Col 3: Declination (equinox 2000),
Col 4: Supernova detection limit, 
Col 5: Color excess from \citet{Schlegel:1998}.
}
\end{deluxetable}

\begin{deluxetable}{lccc|clcc}
\scriptsize
\tablewidth{0pt}
\tablenum{2}
\tablecaption{Thirty eight distant Type Ia Supernovae}
\label{tabsn}
\tablehead{
\colhead{Name} &
\colhead{Redshift} &
\colhead{Discovery} &
\colhead{~~~~} &
\colhead{~~~~~} &
\colhead{Name} &
\colhead{Redshift} &
\colhead{Discovery} \\
\colhead{~} & \colhead{~} & \colhead{$R$ mag.} 
& \colhead{~} & \colhead{~} & 
\colhead{~} & \colhead{~} & \colhead{$R$ mag.}
}
\startdata
1995aq & 0.453 & 22.4 & ~~~~ & ~~~~~ & ~1997ag & 0.592 & 23.2 \cr  
1995ar & 0.497 & 23.1 & ~~~~ & ~~~~~ & ~1997ai & 0.450 & 22.3 \cr
1995as & 0.498 & 23.3 & ~~~~ & ~~~~~ & ~1997aj & 0.581 & 23.8 \cr
1995at & 0.655 & 22.7 & ~~~~ & ~~~~~ & ~1997ak & 0.347 & 24.4 \cr
1995aw & 0.400 & 22.5 & ~~~~ & ~~~~~ & ~1997al & 0.621 & 23.8 \cr
1995ax & 0.615 & 22.6 & ~~~~ & ~~~~~ & ~1997am & 0.416 & 23.4 \cr
1995ay & 0.480 & 22.7 & ~~~~ & ~~~~~ & ~1997ap & 0.830 & 24.2 \cr
1995az & 0.450 & 24.0 & ~~~~ & ~~~~~ & unnamed\tablenotemark{a} 
                                      & $\sim$0.7 & 23.5 \cr
1995ba & 0.388 & 22.6 & ~~~~ & ~~~~~ & ~1997el & 0.636 & 23.1 \cr
1996cf & 0.570 & 22.7 & ~~~~ & ~~~~~ & ~1997em & 0.460 & 23.6 \cr
1996cg & 0.460 & 22.1 & ~~~~ & ~~~~~ & ~1997ep & 0.462 & 22.4 \cr
1996ch & 0.580 & 23.7 & ~~~~ & ~~~~~ & ~1997eq & 0.538 & 22.4 \cr
1996ci & 0.495 & 22.3 & ~~~~ & ~~~~~ & ~1997er & 0.466 & 22.3 \cr
1996ck & 0.656 & 23.5 & ~~~~ & ~~~~~ & ~1997et & 0.633 & 23.4 \cr
1996cl & 0.828 & 23.6 & ~~~~ & ~~~~~ & ~1997eu & 0.592 & 22.4 \cr 
1996cm & 0.450 & 22.7 & ~~~~ & ~~~~~ & ~1997ex & 0.361 & 21.4 \cr
1996cn & 0.430 & 22.6 & ~~~~ & ~~~~~ & ~1997ey & 0.575 & 22.9 \cr
1997ac & 0.320 & 23.1 & ~~~~ & ~~~~~ & ~1997ez & 0.778 & 23.4 \cr
1997af & 0.579 & 23.7 & ~~~~ & ~~~~~ & ~1997fa & 0.498 & 22.5 \cr
\enddata
\tablenotetext{a}{Not observed spectroscopically (see text)}
\tablecomments{
Cols 1 \& 5: IAU Name,
Cols 2 \& 5: Geocentric redshift of supernova or host galaxy,
Cols 3 \& 6: Approximate discovery $R$ magnitude} 
\end{deluxetable}
\tablewidth{0pt}
\begin{deluxetable}{c c c c c c c}
\scriptsize
\tablecaption{SN~Ia rate per unit comoving volume for 
different cosmological models: 
in a flat $\Lambda$ dominated model consistent with the latest 
Supernova Cosmology Project results~(Model~$\Lambda$); 
in a $\Lambda=0$ universe with $\Omega_{\rm M}=0.3$~(Model~O); and in an
Einstein-de-Sitter~universe~(Model~E). 
\label{tabvolfit}}
\tablenum{3}
\tablehead{
Model 
& $\Omega_{\rm M}$
& $\Omega_\Lambda$ 
& $\bar{z}_{\rm exp}\tablenotemark{a}$
& $\bar{z}_{\rm obs}\tablenotemark{b}$ 
& $r_{\rm V}\tablenotemark{c}$ 
& $\alpha\tablenotemark{d}$
} 
\startdata
\multicolumn{7}{l}{\em One parameter Fits}\\
$\Lambda$ & 0.28& 0.72& 0.53 & 0.54 & 1.53$^{+0.28}_{-0.25}$ &\nodata\cr
O         & 0.3 & 0.0 & 0.52 & 0.54 & 2.42$^{+0.44}_{-0.40}$ &\nodata\cr
E         & 1.0 & 0.0 & 0.52 & 0.54 & 3.25$^{+0.58}_{-0.53}$ &\nodata\cr
\multicolumn{7}{l}{\em Two parameters Fits}\\
$\Lambda$ & 0.28 & 0.72 & 0.54 & 0.54 & 1.55$^{+0.29}_{-0.30}$ 
                                      & 0.8$^{+1.6}_{-1.6}$ \cr
O & 0.3 & 0.0 & 0.54 & 0.54 & 2.48$^{+0.48}_{-0.48}$ 
                                      & 1.3$^{+1.6}_{-1.6}$ \cr 
E & 1.0 & 0.0 & 0.54 & 0.54 & 3.36$^{+0.64}_{-0.64}$   
                                      & 1.7$^{+1.5}_{-1.6}$ \cr
\enddata
\tablenotetext{a}{Expected mean redshift. Computed from the expected 
number of supernovae $N_{\rm exp}(z)$ (see text).}
\tablenotetext{b}{Observed mean redshift}
\tablenotetext{c}{Rate per unit volume ($h^3~10^{-4}~\rm{Mpc}^{-3}~\rm{yr}^{-1}$)
at mean redshift $z=\bar{z}_{\rm exp}$. Statistical uncertainty only}
\tablenotetext{e}{Evolution index (see text and Figure 4)}
\end{deluxetable}

\tablewidth{0pt}
\begin{deluxetable}{c c c c c c c}
\scriptsize
\tablecaption{SN~Ia rate per unit luminosity for different 
cosmological models: 
in a flat $\Lambda$ dominated model consistent with the latest 
Supernova Cosmology Project results~(Model~$\Lambda$); 
in a $\Lambda=0$ universe with $\Omega_{\rm M}=0.3$~(Model~O); and in an
Einstein-de-Sitter~universe~(Model~E). 
\label{tablumfit}}
\tablenum{4}
\tablehead{Model 
& $\Omega_{\rm M}$
& $\Omega_\Lambda$ 
& $\bar{z}_{\rm exp}$\tablenotemark{a} 
& $\bar{z}_{\rm obs}$\tablenotemark{b} 
& $r_{\rm L}$\tablenotemark{a} 
} 
\startdata
\multicolumn{6}{l}{\em From CFRS galaxy counts}\\
$\Lambda$ & 0.28 & 0.72 & 0.56 & 0.54 & 0.63$^{+0.11}_{-0.10}$ \\[+0.5ex]
\multicolumn{6}{l}{\em From Luminosity Functions}\\
$\Lambda$ & 0.28 & 0.72 & 0.58 & 0.54 & 0.58$^{+0.10}_{-0.09}$ \\[+0.5ex] 
O & 0.3 & 0.0 & 0.57 & 0.54 & 0.78$^{+0.14}_{-0.13}$ \\[+0.5ex]
E & 1.0 & 0.0 & 0.57 & 0.54 & 0.91$^{+0.16}_{-0.14}$ \\[+0.5ex]
\tableline
\enddata
\tablenotetext{a}{Expected mean redshift. Computed from the expected 
number of supernovae $N_{\rm exp}(z)$ (see text).}
\tablenotetext{b}{Observed mean redshift}
\tablenotetext{c}{Rate per unit luminosity ($h^2~\rm{SNu}$) at mean redshift 
$z=z_{\rm exp}$.
Statistical uncertainty only}
\end{deluxetable}

\tablewidth{0pt}
\begin{deluxetable}{c c c}
\scriptsize
\tablecaption{Summary of Uncertainties 
\label{tabsys}}
\tablenum{5}
\tablehead{
Source 
& $\delta r_{\rm V}$\tablenotemark{a}
& $\delta r_{\rm L}$\tablenotemark{b}
}
\startdata
Cosmological parameters          &$^{+0.25}_{-0.23}$ & $^{+0.04}_{-0.03}$\cr
Detection efficiencies           &$  \pm 0.12 $      &$\pm 0.04$ \cr
Range of Ia lightcurves          &$  \pm 0.14 $      &$\pm 0.05$ \cr 
Field calibration                &$  \pm 0.06 $      &$\pm 0.02$ \cr
Cluster contribution             &$  \pm 0.05 $      &$\pm 0.02$ \cr
Galaxy extinction                &$  \pm 0.02 $      &$\pm 0.01$ \cr
Luminosity estimate              &  \nodata          &$\pm 0.05$ \cr
\tableline
Total syst. uncertainty          &$^{+0.32}_{-0.31}$ & $^{+0.10}_{-0.09}$\cr
\tableline
Statistical uncertainty          &$^{+0.28}_{-0.25}$ & $^{+0.10}_{-0.09}$\cr
\tableline
\enddata
\tablenotetext{a}{Uncertainty on the rate per unit 
volume ($h^3~10^{-4}~\rm{Mpc}^{-3}~\rm{yr}^{-1}$)}
\tablenotetext{b}{Uncertainty on the rate per unit luminosity ($h^2$ SNu)}
\tablecomments{These uncertainties have been computed in a flat
$\Lambda$ dominated universe using 
$\Omega_{\rm M}=0.28^{+0.10}_{-0.09}$ (see text). No estimate was 
made of possible systematic uncertainties
from host galaxy inclination or extinction.}
\end{deluxetable}

\clearpage


\begin{thebibliography}{}

\bibitem[Brinchmann et~al.(1998)Brinchmann et~al.]
{Brinchmann:1998}
Brinchmann, J., Abraham, R., Schade, D., Tresse, L., Ellis, R.S., Lilly,
S., Le Fevre, O., Glazebrook, K., Hammer, F., Colless, M., Crampton, D.,
Broadhurst, T.
1998, \apj, 499, 112

\bibitem[Brinchmann (1999)]
{Brinchmann:1999} 
Brinchmann, J. 
1999, Ph.D. Thesis, University of Cambridge

\bibitem[Burstein \& Heiles(1982)Burstein and Heiles]
{Burstein:1982}
Burstein, D., Heiles, C.
1982, \aj, 87, 1165

\bibitem[Carroll, Press \& Turner(1992)Carroll, Press and Turner]
{Carroll:1992}
Carroll, S.M., Press, W.H., Turner, E.L.
1992, \araa, 30, 499

\bibitem[Cardelli, Clayton \& Mathis(1989)Cardelli, Clayton and Mathis]
{Cardelli:1989}
Cardelli, J. A., Clayton, G. C. Mathis, J. S.
1989, \apj, 345, 245

\bibitem[Cappellaro et~al.(1999)]
{Cappellaro:1999} 
Cappellaro, E., Evans, R., Turatto, M. 
1999, \aa, 351, 459

\bibitem[Cowie, Songaila \& Barger(1999)]
{Cowie:1999} 
Cowie L., Songaila A., Barger A.J. 
1999, \aj, 118, 603

\bibitem[Dahlen \& Fransson(1999)]
{Dahlen:1999} 
Dahlen T., Fransson C.
1999, \aa, 350, 349

\bibitem[Folkes et al.(1999)]
{Folkes:1999} 
Folkes, S., et al. 
1999, \mnras, 308, 459

\bibitem[Garnavich et~al.(1998)Garnavich et~al.]
{Garnavich:1998}
Garnavich P. M., et~al.
1998, \apj, 493, L53

\bibitem[Gibson et~al.(2000)Gibson et~al.]
{Gibson:2000}
Gibson, B. K., et~al. 
2000, \apj, 529, 723 

\bibitem[Gronwall(1995)Gronwall]
{Gronwall:1995}
Gronwall, C.
1995, private communication. 
See also Gronwall, C \& Koo, D, 1995, \apj, 440, L1

\bibitem[Gunn, Hoessel \& Oke(1986)Gunn, Hoessel and Oke]
{Gunn:1986}
Gunn, S., Hoessel, L. \& Oke, B.
1986, \apj, 306, 30

\bibitem[Hamuy {et~al.}(1995)Hamuy {et~al.}]
{Hamuy:1995}
Hamuy, M., Phillips, M.M., Maza, J., Suntzeff, N.B., Schommer, R.,
Aviles, R. 
1995, \aj, 109, 1

\bibitem[Hamuy et~al.(1996)Hamuy et~al.]
{Hamuy:1996}
Hamuy, M., Phillips, M.M., Schommer, R.A., Suntzeff, N.B., Maza, J., 
Aviles, R.
1996, \aj, 112, 2391

\bibitem[Hardin et~al.(2000)]
{Hardin:2000} 
Hardin, D., et al. 
2000, \aap, 362, 419

\bibitem[Hook {et~al.}(2002)Hook {et~al.}]
{Hook:2002}
Hook, I., Nugent, P., et al., SCP Collab,
in preparation

\bibitem[Kim, Goobar \& Perlmutter(1996)Kim, Goobar and Perlmutter]
{Kim:1996}
Kim, A., Goobar, A., Perlmutter, S.
1996, \pasp, 108, 190

\bibitem[Landolt(1992)Landolt]
{Landolt:1992}
Landolt, A. U.
1992, \aj, 104, 340

\bibitem[Leibundgut(1988)Leibundgut]
{Leibundgut:1988}
Leibundgut, B.
1988, Ph. D Thesis, University of Basel

\bibitem[Leibundgut(2001)Leibundgut]
{Leibundgut:2001}
Leibundgut, B.
2001, \araa, 39, 67 

\bibitem[Lilly(1995a)Lilly]
{Lilly:1995a}
Lilly, S.,
1995a, private communication.

\bibitem[Lilly {et~al.}(1995b)Lilly {et~al.}]
{Lilly:1995b}
Lilly, S., Tresse, L., Hammer, F., Crampton, D., Le Fevre, O.
1995b, \apj, 455, 108.

\bibitem[Loveday et al.(1992)]
{Loveday:1992}
Loveday, J. Peterson, B. A., Efstathiou, G. Maddox, S. J.
1992, \apj, 400, L43

\bibitem[Marzke {et~al.}(1998)Marzke {et~al.}]
{Marzke:1998} 
Marzke, R.O., Da Costa, L.N., Pellegrini, P.S., Willmer, C.N.A., Geller, M.J.
1998, \apj, 503, 617

\bibitem[Madau et al.(1998)Madau, Della Valle and Panagia]
{Madau:1998}
Madau, P., Della Valle, M., Panagia, N.
1998, \mnras, 297, L17

\bibitem[Madau \& Pozzetti(2000)]
{Madau:2000} 
Madau P., Pozzetti L. 
2000, \mnras, 312, L9

\bibitem[McMahon \& Irwin(1992)McMahon and Irwin]
{McMahon:1992}
McMahon R. G. and Irwin M. J. 
1992 , Digitized Optical Sky Surveys, eds H. T. MacGillivray \& E. B. 
Thomson, Kluwer p 417.

\bibitem[Nomoto et~al.(1999)]
{Nomoto:1999} 
Nomoto K., Umeda H., Hachisu I., Kato M., Kobayashi C., Tsujimoto T., 
1999, in Truran J., Niemeyer J. eds, `Type Ia Supernovae: Theory and 
Cosmology', Cambridge University Press, astro-ph/9907386

\bibitem[Oke et al.(1995)]
{Oke:1995} 
Oke, J.\ B. et al.\ 
1995, \pasp, 107, 375

\bibitem[Pain {et~al.}(1996)Pain {et~al.}]
{Pain:1996}
Pain, R., et al., SCP Collaboration
1996, \apj, 473, 356

\bibitem[Peacock et~al.(2001)Peacock et~al.]
{Peacock:2001}
Peacock, J. A., et al., 
2001, Nature, 410, 169

\bibitem[Perlmutter {et~al.}(1995a)Perlmutter {et~al.}]
{Perlmutter:1995a}
Perlmutter, S., et~al. SCP Collaboration, 
1995a, in Presentations at the NATO ASI in Aiguablava, Spain, LBL-38400,
  page I.1; also published in Thermonuclear Supernova, P.~Ruiz-Lapuente,
  R.~Canal, and J.Isern, editors, Dordrecht: Kluwer, page 749 (1997)

\bibitem[Perlmutter {et~al.}(1995b)Perlmutter {et~al.}]
{Perlmutter:1995b}
Perlmutter, S., et al., SCP Collaboration
1995b, \iaucirc 6270

\bibitem[Perlmutter {et~al.}(1996)Perlmutter {et~al}]
{Perlmutter:1996}
Perlmutter, S., et al., SCP Collaboration
1996, \iaucirc 6621

\bibitem[Perlmutter {et~al.}(1997a)Perlmutter {et~al.}]
{Perlmutter:1997a}
Perlmutter, S., et~al., SCP Collaboration
1997a, \apj, 483, 565 

\bibitem[Perlmutter {et~al.}(1997b)Perlmutter {et~al.}]
{Perlmutter:1997b}
Perlmutter, S., et al., SCP Collaboration
1997b, \iaucirc 6596

\bibitem[Perlmutter {et~al.}(1997c)Perlmutter {et~al.}]
{Perlmutter:1997c}
Perlmutter, S., et al., SCP Collaboration
1997c, \iaucirc 6804

\bibitem[Perlmutter {et~al.}(1998)Perlmutter {et~al.}]
{Perlmutter:1998}
Perlmutter, S., et~al., SCP Collaboration 
1998, Nature, 391, 51 and erratum (on author list), 392, 311

\bibitem[Perlmutter {et~al.}(1999)Perlmutter {et~al.}]
{Perlmutter:1999}
Perlmutter, S., et al., SCP Collaboration
1999, \apj, 517, 565

\bibitem[Phillips(1993)Phillips]
{Phillips:1993}
Phillips, M.M.
1993, \apjl, 413, L105

\bibitem[Phillips et~al.(1999)Phillips et~al.]
{Phillips:1999}
Phillips, M.M., Lira, P., Suntzeff, N.B., Schommer, R.A., Hamuy, M., Maza., J.
1999, \aj, 103, 1632 

\bibitem[Poggianti(1997)Poggianti]
{Poggianti:1997}
Poggianti B.
1997, A\&AS, 324, 490

\bibitem[Reiss(2002)Reiss]
{Reiss:2002}
Reiss, D.J., 
Private communication, {\it The Rate of Supernovae in the Nearby and Distant Universe}, 
PhD Thesis, University of Washington

\bibitem[Riess, Press \& Kirshner(1995)Riess, Press and Kirshner]
{Riess:1995}
Riess, A. G., Press, W. H. and Kirshner, R. P.
1995, \apj, 438, L17

\bibitem[Riess, Press \& Kirshner(1996)Riess, Press and Kirshner]
{Riess:1996}
Riess, A. G., Press, W. H. and Kirshner, R. P.
1996, \apj, 473, 88

\bibitem[Riess {et~al.}(1998)Riess {et~al.}]
{Riess:1998}
Riess, A. G. et al.
1998, \aj, 116, 1009

\bibitem[Ruiz-Lapuente, Canal \& Burkert(1997)Ruiz-Lapuente, Canal and Burkert]
{Ruiz-Lapuente:1997} 
Ruiz-Lapuente, P., Canal, R., Burkert, A.
1997, in Ruiz-Lapuente, P., Canal, R. Isern, J.
eds, Thermonuclear Supernovae, Kluwer, Dordrecht, p 205

\bibitem[Ruiz-Lapuente \& Canal(1998)Ruiz-Lapuente and Canal]
{Ruiz-Lapuente:1998} 
Ruiz-Lapuente, P., Canal, R., 
1998, \apj 497, L57

\bibitem[Sadat, Guiderdoni \& Blanchard(1998)Sadat, Guiderdoni and Blanchard]
{Sadat:1998}
Sadat, R., Blanchard A., Guiderdoni B., Silk, J.
1998, \aap, 331, L69

\bibitem[Saha et~al.(1999)Saha et~al.]
{Saha:1999}
Saha, A., Sandage, A., Tammann, G.A., Labhardt, L., Macchetto, F.D., 
Panagia, N.
1999, \apj, 522, 802

\bibitem[Schlegel,  Finkbeiner \& Davis(1998)Schlegel, Finkbeiner and Davis]
{Schlegel:1998}
Schlegel, D., Finkbeiner, D., \& Davis, M.
1998, \apj, 550, 525.

\bibitem[Schmidt et~al.(1998)Schmidt et~al.]
{Schmidt:1998} 
Schmidt, B.P., et al., 
1998 \apj, 507, 46

\bibitem[Sullivan et~al.(2000a)Sullivan et~al.]
{Sullivan:2000a} 
Sullivan M., Ellis, R., Nugent, P., Smail I., Madau, P. 
2000a, \mnras, 319, 4549

\bibitem[Sullivan et~al.(2000b)Sullivan et~al.]
{Sullivan:2000b} 
Sullivan M., Treyer M., Ellis R.S., Bridges T., Donas J., Milliard B. 
2000b, \mnras, 312, 442

\bibitem[Sullivan et~al.(2002)Sullivan et~al.]
{Sullivan:2002} 
Sullivan M., Ellis, R.S., et~al., SCP Collab, 
in preparation

%
\bibitem[Yungelson \& Livio(1998)Yungelson and Livio]
{Yungelson:1998}
Yungelson, L., Livio, M.
1998, \apj, 497, 168

\bibitem[Wittman et~al.(1998)Wittman et~al.]
{Wittman:1998}
Wittman, D.\ M., Tyson, J.\ A., Bernstein, G.\ M., Lee, R.\ W., 
dell'Antonio, I.\ P., Fischer, P., Smith, D.\ R., \& Blouke, M.\ M.\ 
1998, SPIE, 3355, 626

\end{thebibliography}

\clearpage

\clearpage

\clearpage
\clearpage

\clearpage

\clearpage

\clearpage

\end{document}